\documentclass[journal]{IEEEtran}
\usepackage{cite}

\ifCLASSINFOpdf
  \usepackage[pdftex]{graphicx}
\else
\fi
\usepackage{amsmath}
\usepackage[mathscr]{euscript}
\usepackage{algorithmic}
\usepackage[linesnumbered, ruled]{algorithm2e}
\usepackage{setspace}
\usepackage{multirow}
\usepackage{amsthm}
\usepackage{amsfonts}
\usepackage{hyperref}
\usepackage{xcolor}

\usepackage{array}
\usepackage{url}
\usepackage{xcolor}
\newcounter{tempEquationCounter}
\newcounter{thisEquationNumber}

\newcounter{tempAlgorithmCounter}
\newcounter{thisAlgorithmNumber}

\begin{document}
%
\title{A Diffeomorphic Flow-based Variational Framework for Multi-speaker Emotion Conversion}

%

\author{Ravi~Shankar,
        Hsi-Wei~Hsieh, 
        Nicolas~Charon, 
        and~Archana~Venkataraman,~\IEEEmembership{Member,~IEEE,}
\thanks{Ravi Shankar and Archana Venkataraman are with the Department
of Electrical and Computer Engineering, Johns Hopkins University, Baltimore,
MD, 21218 USA e-mail: \{rshanka3, archana.venkataraman\}@jhu.edu. Hsi-Wei Hsieh and Nicolas Charon are with the Department of Applied Math and Statistics, Johns Hopkins University, Baltimore, Md, 21218 USA email:\{hhsieh, charon\}@cis.jhu.edu. This paper has supplementary downloadable material available at \url{https://engineering.jhu.edu/nsa/wp-content/uploads/2022/07/demo_VCGAN_CGAN.zip}}
}
%
%

\markboth{Journal of \LaTeX\ Class Files,~Vol.~20, No.~14, August~2021}%
{Shell \MakeLowercase{\textit{et al.}}: Bare Demo of IEEEtran.cls for IEEE Journals}
%



\maketitle

\begin{abstract}
This paper introduces a new framework for non-parallel emotion conversion in speech. 
Our framework is based on two key contributions. 
First, we propose a stochastic version of the popular Cycle-GAN model. 
Our modified loss function introduces a Kullback–Leibler (KL) divergence term that aligns the source and target \textit{data distributions} learned by the generators, thus overcoming the limitations of sample-wise generation. 
By using a variational approximation to this stochastic loss function, we show that our KL divergence term can be implemented via a paired density discriminator. 
We term this new architecture a variational Cycle-GAN (VCGAN). 
Second, we model the prosodic features of target emotion as a smooth and learnable deformation of the source prosodic features. 
This approach provides implicit regularization that offers key advantages in terms of better range alignment to unseen and out-of-distribution speakers. 
We conduct rigorous experiments and comparative studies to demonstrate that our proposed framework is fairly robust with high performance against several state-of-the-art baselines. 
\end{abstract}


%
\IEEEpeerreviewmaketitle

\section{Introduction}
\IEEEPARstart{S}{peech} is perhaps our primary mode of communication as humans. 
It is a rich medium, in the sense that both semantic information and speaker intent are intertwined together in a complex manner. 
The ability to convey emotion is an important yet poorly understood attribute of speech. 
Common work in speech analysis focuses on decomposing the signal into compact representations and probing their relative importance in imparting one emotion versus another. 
These representations can be broadly categorized into two groups: acoustic features and prosodic features. 
Acoustic features (e.g., spectrum) control resonance and speaker identity. 
Prosodic features (e.g., F0, energy contour) are linked to vocal inflections that include the relative pitch, duration, and intensity of each phoneme. 
Together, the prosodic features encode stress, intonation, and rhythm, all of which impact emotion perception. 
For example, expressions of anger often exhibit large variations in pitch, coupled with increases in both articulation rate and signal energy. 
In this paper, we develop an automated framework to transform an utterance from one emotional class to another. 
The problem, known as \textit{emotion conversion}, is an important stepping stone to affective speech synthesis.

Broadly, the goal of emotion conversion is to modify the perceived affect of a speech utterance without changing its linguistic content or speaker identity. 
This setting allows the user to control the speaking style, while allowing the model to be trained on limited data. 
Emotion conversion is a particularly challenging problem due to the inherent ambiguity of emotions themselves~\cite{face_vocal_expression, psych}. 
The boundaries between emotion classes are also blurry, and prior knowledge about the speaker can sometimes play a major role in the emotion perception. 
That being said, one of the main application of emotion conversion is to evaluate the quality of human-machine dialog systems~\cite{prosody_eval}. 
Here, intonation changes can indicate the level of naturalness of a conversation between a machine and a person. 
Emotion conversion can also be helpful in studying neurodevelopmental disorders such as autism, which is characterized by poor emotion perception capability. 
On the technical front, being able to control the granularity of the emotion expression in synthesized speech is an important step towards developing an intelligent conversational system. 
Finally, emotion conversion can be used for data augmentation when training emotion classification or speaker recognition systems~\cite{stargan_original, emotion_stargan}. 

Early work in emotion conversion traces its roots to neuroscientific experiments, which were designed to study the influence of emotions in the brain. 
Interestingly, many of the implicated features tend to generalize across languages. 
For example, the work of~\cite{emotion_perception_factors} determined the F0 (i.e., pitch) contour and the energy (loudness) profile as the main factors responsible for primary emotions. 
Additionally, voice quality and utterance duration have also been identified as features affecting emotion perception~\cite{zeynep_independent_model}. 
Voice quality is a function of the spectrum representation and duration can be called as a proxy for the speaking rate. 
A comprehensive study was conducted by~\cite{japanese_emotion} to understand the impact of systematically changing acoustic and prosodic features on emotional perception. 
These experiments were performed on a Japanese language database with some consistency shown for English. 

Algorithms for emotion conversion fall into three general categories. 
The first approach relies on constructing a statistical model of the source and target prosodic features to allow inference from one domain to another. 
One example of this approach is the work of~\cite{cart_emotion_conversion}, which uses classification and regression trees (CART) to modify the F0 contour in Mandarin. 
An alternate strategy uses a Gaussian Mixture Model (GMM) for voice and emotion conversion. 
The central idea is to learn a GMM that captures the joint distribution of the source and target emotional speech features during training. 
Inference of a new conversion is done via the conditional mean of the target features given the test source features. 
Mathematically, let $\mathbf{z}_i = [\mathbf{x}_i \quad \mathbf{y}_i]^{\mathrm{T}}$ denote the concatenated source and target features and $c_i$ denote the latent cluster assignment for utterance~$i$. 
From here, we have:
\begin{equation}
    P(\mathbf{z}_i|c_i) = \sum_{k=1}^K P(\mathbf{z}_i | c_i=k) P(c_i = k)
    \label{eqn:GMM_basic}
\end{equation}
where, $P(\mathbf{z}_i|c_i=k) \sim N(\mathbf{z}_i; \boldsymbol{\mu}_k, \boldsymbol{\Sigma}_k)$ and its parameters are estimated via the Expectation-Maximization (EM) algorithm, along with the latent prior $P(c_i=k)$.

Using properties of the Gaussian distribution, it can be shown that the conditional mean of the target features $\mathbf{y}_i$ given the source features $\mathbf{x}_i$ is given by the expression
\begin{align}
    E[\mathbf{y}_i|\mathbf{x}_i] = \sum_{k=1}^K P(c_i = k | \mathbf{x}_i) \Big[ \boldsymbol{\mu}_k^y + \boldsymbol{\Sigma}_k^{xy} (\boldsymbol{\Sigma}_k^{xx})^{-1} (\mathbf{x}_i - \boldsymbol{\mu}_k^x) \Big] 
    \label{eqn:GMM_inference}
\end{align}
where, $P(c_k | x)$ can be computed via Bayes' Rule. 
One of the main drawback of this approach is the over-smoothing of spectral parameters in inference stage due to averaging effect. 
To counter this, a global variance constraint based inference proposed by~\cite{global_variance_GMM} was adopted for emotion conversion by~\cite{gmm_emo_conv}. 


The second approach for emotion conversion is based on sparse recovery~\cite{nmf_emo_conv}. 
This technique entails learning an over-complete dictionary of both acoustic and prosodic features for each emotion class. 
During conversion, the input utterance is first decomposed using the source emotion dictionary by estimating a coefficient matrix with sparsity prior. 
These coefficients are then used for reconstruction using the target emotion dictionary elements/atoms. 
The authors of~\cite{nmf_emo_conv} used active Newton-set~\cite{active_set_newton} based non-negative matrix factorization~\cite{nmf_paper} to estimate the sparse coding. 
Mathematically, given a non-negative matrix input $\mathbf{X}$ (e.g., spectrogram magnitude), we seek non-negative matrices $\mathbf{U}$ and $\mathbf{V}$ to minimize:
\begin{equation}
    \mathcal{J} = \Arrowvert \mathbf{X} - \mathbf{U}\mathbf{V} \Arrowvert_F^2 + \lambda \sum_j \Arrowvert \mathbf{V}(:,j) \Arrowvert_1
    \label{eqn:sparse_nmf}
\end{equation}
The first term in Eq.~(\ref{eqn:sparse_nmf}) enforces the data fidelity, whereas the second term encourages sparsity of the learned encoding $\mathbf{V}$. The variable $\mathbf{U}$ denotes the overcomplete dictionary.

The third approach for emotion conversion relies on deep neural networks to automatically learn complex and nonlinear speech modifications. 
For example, a bidirectional LSTM approach has been suggested by~\cite{bi_lstm, lstm_emo_conv} for modifying the prosodic features. 
The authors further proposed using a continuous wavelet transform based parameterization for the F0 and energy contour to decompose into segmental and supra-segmental components. 
Our prior work proposed an alternative method for prosodic modification based on highway neural networks~\cite{highway_net, hnet_max_likelihood}, which maximize the representation log likelihood in an EM algorithm setting. 
We further proposed an F0 modification scheme using the principle of diffeomorphic curve warping as a smoothness prior for the transformed F0 contour~\cite{diffeomorphic_hnet}. 
This diffeomorphic parameterization was extended to spectrum modification in~\cite{chained_model}. 
Specifically, we used a latent variable regularization technique to sequentially modify the F0 contour and the spectrum.

The methods discussed so far belong to the domain of supervised learning. 
Namely, they rely on labeled parallel speech data to learn the requisite emotion conversion. 
Curating parallel corpora is expensive, which explains why there are only a handful of such databases~\cite{vesus} available online. 
Beyond data scarcity, most supervised emotion conversion methods require the parallel utterances to be time-aligned using dynamic time warping (DTW)~\cite{dtw} prior to analysis. 
This alignment procedure allows us to learn a frame-wise mapping between the source and target utterances. 
While simple and apt for smaller corpora, DTW is prone to errors, particularly during periods of silence or unvoiced sounds.

The current iteration of methods focus on unsupervised emotion conversion and do not require parallel data. 
These models rely on expressiveness of neural networks to learn a parametric distribution for each pair of emotions. 
One of the most prominent model in this space is Generative Adversarial Network (GAN). 
Mathematically, let $G$ and $D$ denote the generator and discriminator, respectively. 
The objective of the GAN is a minimax loss given by the following:
\begin{align} \nonumber
    \mathcal{L}_{adv} = \min_G \max_D \; & E_{x \sim P(X)}[\log(D(x)] \\ \label{eqn:GAN_objective}
    & + E_{z \sim P(Z)}[\log(1 - D(G(z))]
\end{align}
where $P(X)$ denotes the data distribution and $P(Z)$ denotes a noise density which is usually Normal i.e, $N(0,I)$. 

The Cycle-GAN architecture goes one step beyond Eq.~(\ref{eqn:GAN_objective}) by tying two separate GANs together via a cycle consistency objective. 
Formally, let $A$ and $B$ denote the domains of the source and target data distributions. 
The two generators in Cycle-GAN are tasked with learning transformation from $A\rightarrow B$ and $B \rightarrow A$, respectively. 
The cycle consistency loss connects the generators by enforcing that the sequence of transformations, i.e. $A \rightarrow B \rightarrow A$ should look similar to the original input. 
For clarity, we will refer to these generators as the ``forward" and ``backward" transformations of the Cycle-GAN and use the notation $G_{\gamma}$ (forward) and $G_{\theta}$ (backward). 
Mathematically, the cyclic objective is written as:
\begin{equation}
\mathscr{L}_{cycle} = E_{x \sim P(X)} \left[\lVert x - G_{\theta}(G_{\gamma}(x)) \rVert_{1}\right]
\label{eqn:cycle_loss}
\end{equation}

The algorithm of~\cite{cyclegan_emo_conv_2} uses a Cycle-GAN to disentangle the content and style of a speech utterance into two separate variables based on \textit{a priori} information embedded into the network architecture. 
Another approach proposed by~\cite{cyclegan_emo_conv} uses a Cycle-GAN to transform the F0 contour and spectrum, as parameterized by a discrete wavelet transform, for emotion conversion. 
A Star GAN~\cite{stargan_original} model proposed by~\cite{emotion_stargan} relies on a multi-task discriminator and a single generator for conversion between multiple emotional classes. 
Due to the poor quality of generated samples, the authors used this method for data augmentation to improve emotion classification accuracy, rather than for speech synthesis. 
While all these methods show tremendous promise, one common drawback is that they have been trained and evaluated on single speaker datasets. 
Thus, it is unclear how they will perform in either a multi-speaker or an out-of-sample generalization setting.

In this paper we propose a novel technique for emotion conversion using a variational formulation of the Cycle-GAN.
Our novel loss formulation leads to a joint density discriminator which minimizes the upper bound on KL-divergence between the target data density and its parameterized counterpart. 
Our method further learns the target emotion F0 and energy contour by modeling them as a smooth deformation of the source emotion features. 
A preliminary version of this work appeared in Interspeech 2020~\cite{vcgan}.
This paper provides the following novel contributions above the conference paper. 
First, we model the transformation of F0 and energy contours of an utterance jointly using intermediate hidden variables.
This is in contrast with the previous approach where we modify the F0 contour and spectrum, independently. 
Second, our graphical model for the conversion strategy allows us to disentangle the discriminator's objective for energy and F0 contour using conditional independencies directly inferred from the graph. 
Finally, we evaluate our proposed framework in both, a multi-speaker setting as well as on out-of-distribution speakers which the model does not see during training. 
We further provide comparative studies about the distribution and stability properties of our technique with a state-of-the-art baseline.


\begin{figure}[!t]
  \centering
  \includegraphics[width=0.9\linewidth, height=4.5cm]{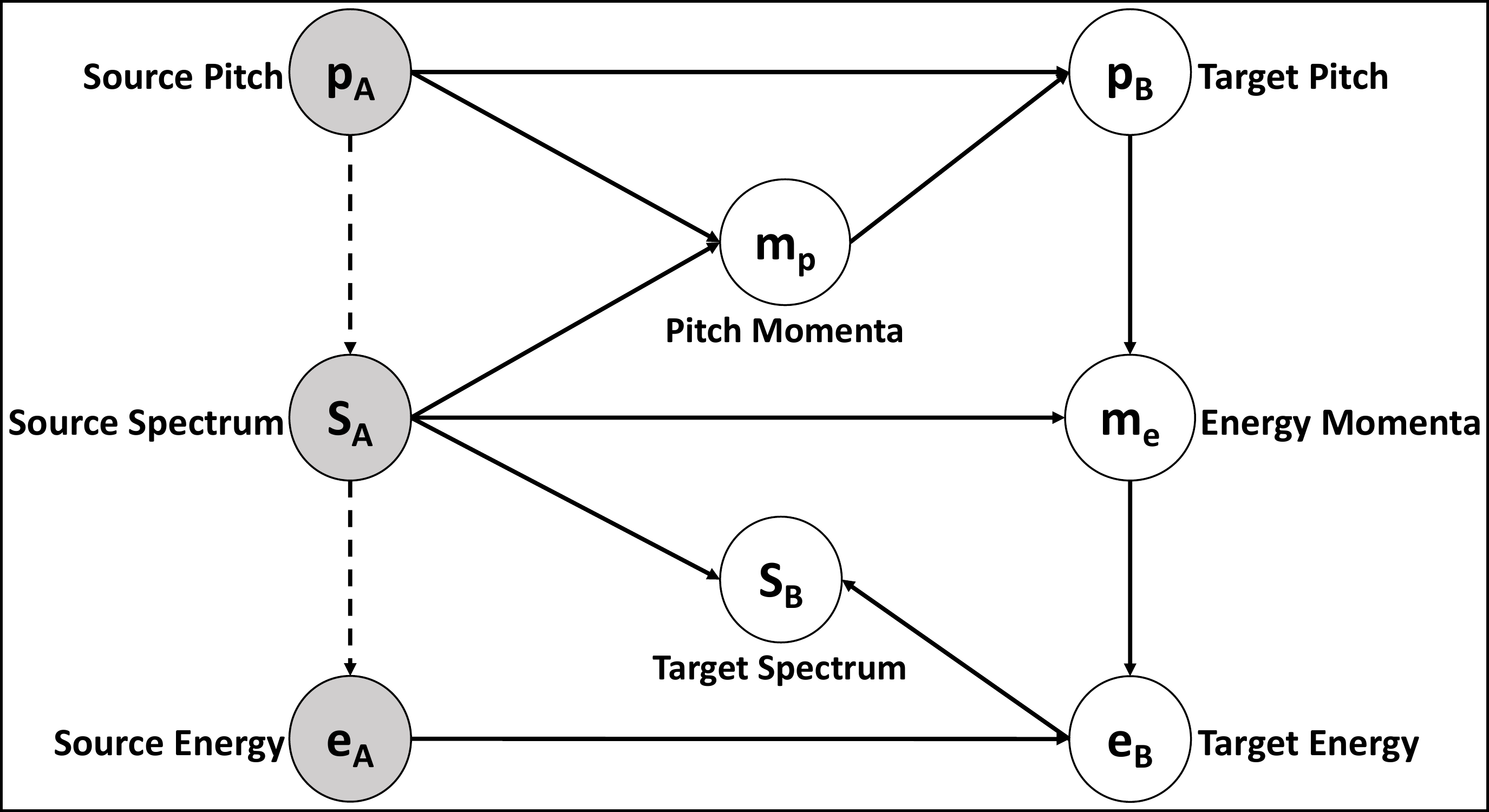}
  \caption{Graphical representation of our emotion conversion strategy. $\mathbf{m_p}$ and $\mathbf{m_e}$ serve as an intermediaries for pitch and energy contours, respectively.}
  \label{fig:graphical_model}
\vspace{-5mm}
\end{figure}

\begin{figure*}[t]
  \centering
  \includegraphics[width=0.95\textwidth, height=4.5cm]{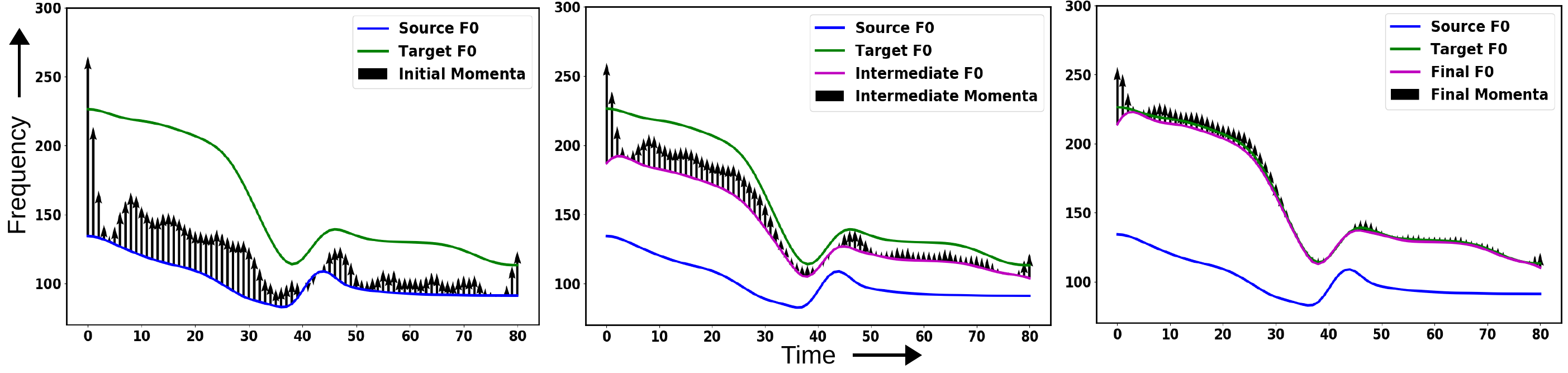}
  \caption{Demonstration of the warping procedure. The leftmost figure shows the source and target F0 contours and the initial momenta. The middle figure shows the F0 contours at an intermediate time step. The rightmost figure is the final result of warping where modified F0 contour matches with the target F0.}
  \label{fig:momenta_demo}
\end{figure*}

\section{Method}
Our strategy is to manipulate two key prosodic features: the F0 (pitch) contour, and the energy (loudness) contour. 
Fig.~\ref{fig:graphical_model} shows the relationship between the features during the inference step of the process. 
We begin with taking an utterance in source emotion A from which we extract the F0 contour $(\mathbf{p}_A)$ and the mel-cepstral features $(\mathbf{S}_A)$ using the WORLD vocoder~\cite{world_vocoder}. 
The energy contour $(\mathbf{e}_A)$ is extracted directly from the spectral features. 
We define latent variables called momenta $(\mathbf{m}_p, \mathbf{m}_e)$, which serve as intermediaries between the two emotion classes under consideration. 
The F0 contour in target emotion $(\mathbf{p}_B)$ is a deterministic function of the momenta $(\mathbf{m}_p)$ and source F0 contour, through a diffeomorphic warping process that we describe in Section~\ref{diffeo_sub}. 
The estimated F0 contour and the source spectrum together generate the momenta $(\mathbf{m}_e)$ for the energy contour which is then further used to generate the cepstral features $(\mathbf{S}_B)$. 
The estimated F0 contour and cepstral features combine together to give the converted utterance in the target emotion B. 

We take an unsupervised approach to model training and evaluation using a Cycle-GAN framework. This strategy allows us to handle non-parallel and multi-speaker datasets. For robustness, we introduce a novel KL-divergence loss to align the \textit{distribution} of the source and target emotional classes, as described in Section~\ref{sec:VCGAN} . The KL-divergence gives rise to a new class of discriminators that operate on pairs of samples.

To summarize, our technical innovations are as follows: 
\begin{itemize}
    \item We propose a joint model for F0 and energy modification which uses latent variables called momenta as an intermediary between source and target emotion features.
    \item We highlight several shortcomings of cyclic consistency loss which is the backbone of our baseline reference model and analyze them theoretically.
    \item We propose a new KL-divergence penalty and minimize its upper bound to address the limitations of cyclic loss. We verify its advantages through multiple experiments.
    \item We evaluate our model on multiple experiment paradigms i.e, single speaker, mixed speaker, leave-one-fold and Wavenet to paint a complete picture of our model.
\end{itemize}

\subsection{Variational Cycle-GAN} \label{sec:VCGAN}
The cycle consistency loss of a traditional Cycle-GAN is given by Eq.~(\ref{eqn:cycle_loss}) and repeated below for convenience:
\begin{equation} \label{eqn:cycle}
    \mathscr{L}_{cycle} = E_{x \sim P(X)} \left[\lVert x - G_{\theta}(G_{\gamma}(x)) \rVert_{1}\right]
\end{equation}
This formulation imposes just a point-wise regularization on the input $X$ and the cyclic converted sample~$G_\theta(G_\gamma(x))$.

It is easy to show that Eq.~(\ref{eqn:cycle}) is not a well-behaved loss function (Propositions 1~and~2 in Appendix~A). Specifically,
\begin{enumerate}
    \item It only enforces a first-order moment matching between the generated and target data distributions.
    \item The expectation in Eq.~(\ref{eqn:cycle}) depends on the sampling variance, which leads to a noisy gradient estimate when optimizing the parameters of the generator.
\end{enumerate}

The first point establishes a weak coupling between the two generators. 
In addition, the discriminators~$D_{\theta}$ and~$D_{\gamma}$ do not have information about the complementary generators when training a traditional Cycle-GAN. 
At a high level, the min-max game played by the generators and discriminators is operating on incomplete information about the underlying data.

The second point often results in poor calibration of the gradients under scenarios where the target distribution is perfectly learnable. 
Practically speaking, this sampling variance is unknown, which can lead to instability during the optimization. For example, it may prompt the generator to take a step that does not reduce the cycle consistency loss (e.g., overshooting the local optimum). Further, because this variance is inherently tied to the parameters of the neural network, the generators can potentially end up learning a null or an identity function in order to minimize the expected cycle consistency loss (e.g., mode collapse). Finally, due to the expected loss being a function of the dimensionality of the data, it scales the gradients computed during backpropagation making the impact of sampling variance more pronounce.

We approach these problems by considering KL-divergence based penalty on the input data distribution and the cyclic transformation. 
Formally, let $(\mathbf{S}_A, \mathbf{p}_A)$ and $(\mathbf{S}_B, \mathbf{p}_B)$ be the source and target cepstrum and F0 contours of two \textit{non-parallel} utterances in emotion A and B, respectively. 
The generators are denoted by $G_{\gamma} : (\mathbf{S}_A, \mathbf{p}_A) \rightarrow (\mathbf{S}_B, \mathbf{p}_B)$ and $G_{\theta} : (\mathbf{S}_B, \mathbf{p}_B) \rightarrow (\mathbf{S}_A, \mathbf{p}_A)$. 
The corresponding distributions learned by the generator functions are given by $P_{\gamma}(\mathbf{S}_B, \mathbf{p}_B)$ and $P_{\theta}(\mathbf{S}_A, \mathbf{p}_A)$. Our new penalty for the generator $G_{\gamma}$ is:
\begin{equation}
\mathscr{L}_{G_\gamma} = KL\Big(P(\mathbf{S}_A, \mathbf{p}_A) \Arrowvert P_{\theta}(\mathbf{S}_A, \mathbf{p}_A)\Big)
\label{eqn:KL_loss}
\end{equation}

Using the law of total probability, we can write:
\begin{align} \nonumber
    P_{\theta}(\mathbf{S}_A, \mathbf{p}_A) = \int \int &P_{\theta}(\mathbf{S}_A, \mathbf{p}_A | \mathbf{S}_B, \mathbf{p}_B) \\ \label{eqn:total_probability}  
    &\times P(\mathbf{S}_B, \mathbf{p}_B) \; d\mathbf{S}_B \; d\mathbf{p}_B
\end{align}

Eq.~(\ref{eqn:total_probability}) is generally intractable, but we can derive an upper bound on the loss in Eq.~(\ref{eqn:KL_loss}) that can be optimized easily (see Appendix~B). 
Effectively, we can minimize: 
\begin{align} \nonumber
    \tilde{\mathscr{L}}_{G_{\gamma}} = E_{(\mathbf{S}_A, \mathbf{p}_A)} &\Big[ E_{(\mathbf{S}_B, \mathbf{p}_B) \sim P_{\gamma}} \Big[ \log \Big( P_{\gamma}(\mathbf{S}_B, \mathbf{p}_B | \mathbf{S}_A, \mathbf{p}_A) \\ \label{eqn:KL_upper_bound}
    &\times P(\mathbf{S}_A, \mathbf{p}_A) \Big) \Big] \Big]    
\end{align}

Eq.~(\ref{eqn:KL_upper_bound}) highlights an important difference between traditional Cycle-GAN and our variational approach. 
Namely, our min-max objective leverages higher-order relationships by comparing the joint density of source and target data factorized by the two generators. 
This transparency is noticeably absent in the traditional Cycle-GAN, in which the discriminator operates on the marginal densities $P(\mathbf{S}_A, \mathbf{p}_A)$ and $P_{\theta}(\mathbf{S}_A, \mathbf{p}_A)$ to determine whether the sample is ``real" or ``fake". 
Finally, we implement the spectrum modification module solely by changing the energy contour; this strategy avoids degradation in speech quality due to errors in spectrum prediction. We have conducted an experiment (see Appendix~G), which demonstrates no difference in user preference for speech generated with the original (mismatched) spectrum and speech generated with a modified spectrum based on the new F0 contour.

\subsection{Prosodic Regularization via Momenta}\label{diffeo_sub}
As shown in the Fig.~\ref{fig:graphical_model}, we use two intermediate representations (denoted by $\mathbf{m_p}$ and $\mathbf{m_e}$) to model the transition of prosodic features from the source to target emotion. 
This technique can be viewed as an implicit regularization on the conversion procedure. 
Practically, we model the target prosodic contours as a smooth deformation of the source F0/energy contours. 
This idea stems from the domain of image registration where a moving image is iteratively deformed to align or match with a fixed image~\cite{image_registration}. 
We adapt this registration framework from 2-dimensional image surfaces to 1-dimensional curves in the Euclidean space. 

While there are multiple ways to represent the deformation process, one popular technique is known as the Large Deformation Diffeomorphic Metric Mapping (LDDMM)~\cite{diffeomorphism, landmark_diffeomorphism}. 
These functions are defined as a smooth and invertible mapping between two topological manifolds. 
An important feature of this LDDMM model is the ability to parameterize diffeomorphic transformations by low-dimensional embeddings known as momenta~\cite{momenta_control_diffeomorphism}. 
Effectively, the source prosodic contour specifies the initial state, while the momenta ($\mathbf{m_p}$) specifies the initial trajectory of the dynamical system. 
Thus, specifying the input curve and momenta are sufficient to generate the  final state of a target curve. 
Fig.~\ref{fig:momenta_demo} shows an example of momenta acting on a source F0 contour to match it with a target F0 contour via the LDDMM registration process.

Mathematically, let \textbf{$\mathbf{p}_{A}^{\mathbf{t}}$} and \textbf{$\mathbf{p}_{B}^{\mathbf{t}}$} denote a pair of source and target F0 contours, respectively. 
The variable $\mathbf{t}$ corresponds to the location of the analysis window as it moves across a given speech utterance. 
The goal of the deformation process is to estimate a series of small vertical displacements $\mathbf{v_{t}(x;s)}$ over frequency and time. 
The integral of these small displacements produces a final large vector field denoted by $\phi_{\mathbf{t}}^{\mathbf{v}} = \int_{0}^{1}\mathbf{v}_{\mathbf{t}}(\cdot;s)ds$~\cite{diffeomorphism}. 
Representing the momenta variable by $\mathbf{m}_p$, the LDDMM objective function can be written as:

{\small
\begin{equation}
    \Gamma(\mathbf{m}_p) = \frac{1}{2} 
    \sum_{i,j=1}^{T}\gamma_{ij}[\mathbf{m}_p]_i [\mathbf{m}_p]_j + \lambda \sum_{t=1}^{T} \Arrowvert \phi_{\mathbf{t}}^{\mathbf{v}}(\mathbf{p}_{A}^{\mathbf{t}};1) - \mathbf{p}_{B}^{\mathbf{t}} \Arrowvert_{2}^{2}
    \label{eqn:momenta_objective}
\end{equation}}
The variable $\gamma_{ij}$ is an exponential smoothing kernel evaluated on pairs of time points of the source contour $\mathbf{p}_{A}^{\mathbf{t}}$ whereas, $\lambda$ is the trade-off between smoothness of momenta and the difference between the source and target F0 contours. 

Rather than solving Eq.~(\ref{eqn:momenta_objective}) explicitly to obtain the momenta, we estimate it blindly via sampling from the generators. 
From a practical standpoint, the continuous time process specified by LDDMM can be easily discretized to run for a fixed number of iterations. 
The main advantage of using a latent regularizer is that it allows the F0 and energy contours to be generated in a dynamically controlled fashion. 
Adversarial training can be susceptible to mode collapse due to imbalance between generator-discriminator losses, learning rates, and the architecture of the neural networks. 
Deformation based F0 estimation stabilizes the generative process and prevents it from swinging wildly and leading to mode collapse. 
We will also demonstrate that this latent regularization improves the generalization capabilities of our framework to unseen speakers. 
Algorithm~\ref{alg:momenta_to_pitch} outlines the warping process given a momenta, an F0 contour and an exponential smoothing kernel having a scale $\sigma$. 
This scale parameter controls the smoothness of the velocity vector fields and is fixed for all our experiments.

\begin{figure*}[t]
  \centering
  \includegraphics[width=0.95\textwidth, height=8.3cm]{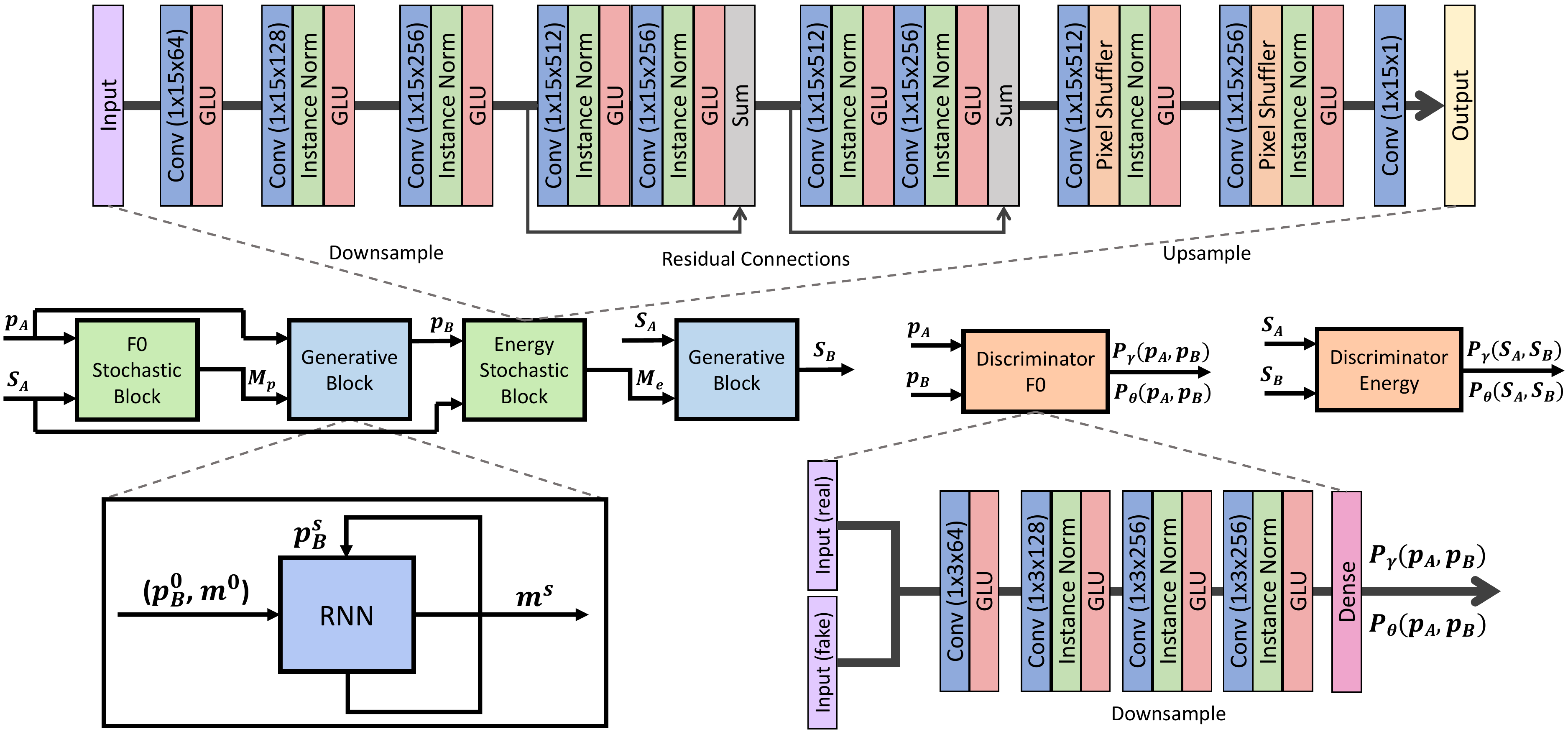}
  \caption{Architecture of the neural network for F0 and energy prediction. The output of F0 prediction is fed as input for energy estimation. Each generator has two blocks: a stochastic block for sampling momenta and a generative/deterministic block for curve warping (represented as an RNN).}
  \label{fig:network_architecture}
\end{figure*}

\subsection{Hybrid Generative Architecture}
Our F0/energy conversion is a two-step process: first, we estimate the momenta, then, we modify the source prosodic contours via a deterministic warping using momenta. 
Our generators mimic this process by integrating a stochastic component with trainable parameters and a deterministic component with fixed/static parameters. 
The stochastic component for F0 momenta prediction takes the spectrum and source F0 as its inputs. 
For energy momenta prediction, the stochastic component relies on the source spectrum (which implicitly contains the energy information) and converted F0. The dimensions of the momenta are the same as F0 and energy contour. 
We empirically fix the smoothness parameter, $\sigma$ at 50 for F0 and at 2 for energy contour to span the appropriate ranges. 
We adapt the 1-D convolutional architecture from~\cite{cyclegan_vc} for the stochastic block of the generators as shown in Fig.~\ref{fig:network_architecture}. 
It has been experimentally verified that fully convolutional networks are more stable in a GAN framework than including fully-connected layers~\cite{dcgan}. 
The deterministic LDDMM warping function can be represented as a recurrent neural network (RNN) with a fixed set of parameters due to its iterative nature. 

We constrain the generators to sample smoothly varying momenta by adding a Laplacian penalty $\mathscr{L}_{m} = E[ \Arrowvert \nabla \mathbf{m}_{p}\Arrowvert^2 ] + E[ \Arrowvert \nabla \mathbf{m}_{e}\Arrowvert^2 ]$ to the overall generator loss. 
The gradient of this term is approximated by the first-order difference of the momenta along the time axis. 
The final objective to minimize for the loss of generator $G_{\gamma}$ is as follows:

{\footnotesize
\begin{align} \nonumber
    &\mathscr{L}_{G_{\gamma}} = \lambda_{c_1} E \Big[ \Arrowvert \mathbf{p}_{A} - \mathbf{p}_{A}^{c} \Arrowvert \Big] + \lambda_{m} E \Big[ \Arrowvert \nabla \mathbf{m}_{p} \Arrowvert_2^2 + \Arrowvert \nabla \mathbf{m}_{e} \Arrowvert_2^2 \Big] \\ \nonumber
    & + \lambda_i E \Big[ \Arrowvert \mathbf{e}_A - \mathbf{e}_A^I \Arrowvert \Big] + \lambda_{c_2} E \Big[ \Arrowvert \mathbf{e}_A - \mathbf{e}_A^c \Arrowvert \Big] \\ \label{eqn:generator_loss_final}
    & + \lambda_{d} E_{(\mathbf{S}_{A},\mathbf{p}_{A})} \Big[E_{(\mathbf{S}_B, \mathbf{p}_{B}) \sim P_{\gamma}}  \Big[ \log \Big( D_{\gamma}(\mathbf{S}_{A}, \mathbf{p}_{A}, \mathbf{S}_{B}, \mathbf{p}_{B}) \Big) \Big] \Big]
\end{align}}

In the case of energy contour modification, we add an identity loss to the generator, which keeps the modified contour ``close" to the original. 
The superscripts $I$ and $c$ denote the identity and cyclic components, respectively. 
Identity loss has been proposed by~\cite{cycle_gan} in Cycle-GANs to make the generators more robust and allow them to reduce distortion when presented with a sample from target density itself. 
We omit the identity loss for the F0 conversion, as this contour tends to vary widely across utterances and emotional classes. 

Finally, we update the parameters of the stochastic block of the generators by back-propagating through the deterministic LDDMM transformation, as implemented by an RNN.


\begin{algorithm}[!t]
\setstretch{1.2}
    \SetKwInOut{Input}{Input}
    \SetKwInOut{Output}{Output}
    function \underline{GenerateF0} $(\mathbf{m}_p,\mathbf{p}_A)$\;
    \Input{momenta ($\mathbf{m}_p$) and source F0 ($\mathbf{p}_A$)}
    \Output{target F0 ($\mathbf{p}_B$)}
    Set $s = 0$, $[\mathbf{p}_B]^0 = \mathbf{p}_A$ and $[\mathbf{m}_p]^0 = \mathbf{m}_p$\;
    \eIf{$ s < 5 $}
      {
        $d_{i,j} \leftarrow [\mathbf{p}_A]_i^s - [\mathbf{p}_A]_j^s $\;
        $K_{i,j} \leftarrow \exp{-\frac{(d_{i,j})^2}{\sigma^2}}$\;
        $[\mathbf{p}_B]_{i}^{s+1} \leftarrow [\mathbf{p}_B]_{i}^{s} + \sum_{l}K_{i,l} \cdot [\mathbf{m}_p]_{l}^{s}$\;
        $[\mathbf{m}_p]_{i}^{s+1} \leftarrow [\mathbf{m}_p]_{i}^{s} + 2 \cdot \sum_{j} \frac{-K}{\sigma^2} \; d_{i,j} \cdot [\mathbf{m}_p]_{i}^{s} \; [\mathbf{m}_p]_{j}^{s}$\;
        $s \leftarrow s+1$\;
      }
      {
        return $[\mathbf{p}_B]^{s}$\;
      }
    \caption{Warping to generate the target F0 contour given the momenta and source F0 contour}
    \label{alg:momenta_to_pitch}
\end{algorithm}

\begin{figure*}[t]
  \centering
  \includegraphics[width=0.99\textwidth, height=4.7cm]{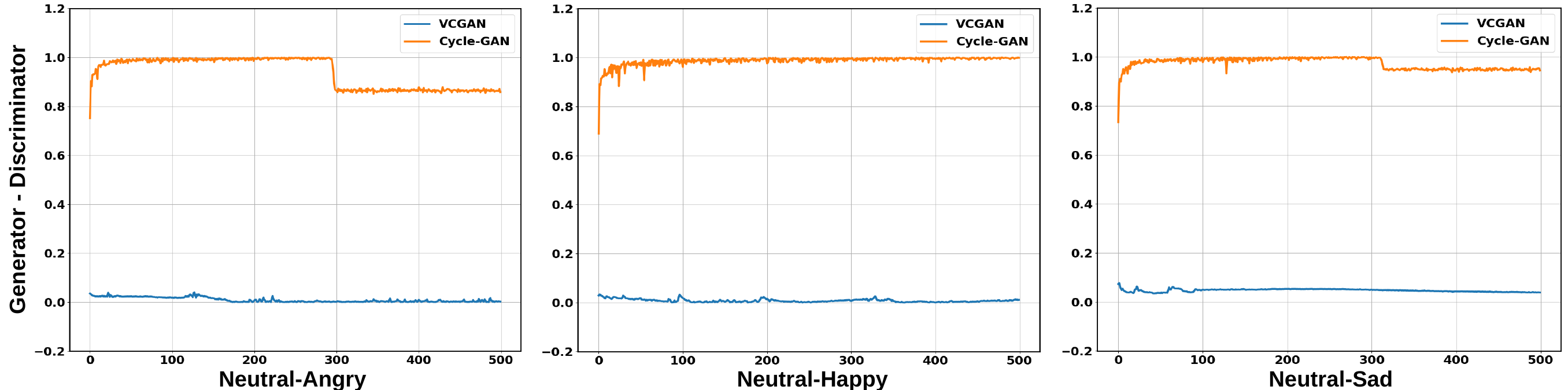}
  \caption{Comparing Cycle-GAN with its variational counterpart. On Y-axis, we denote the difference between the generator and discriminator loss. On X-axis, we denote the number of epochs. The plots represent the mismatch between the adversarial losses which is an indicator of instability in training~\cite{improving_gans}.}
  \label{fig:stability_result}
\end{figure*}

\begin{figure}[!t]
  \centering
  \includegraphics[width=0.96\linewidth, height=11cm]{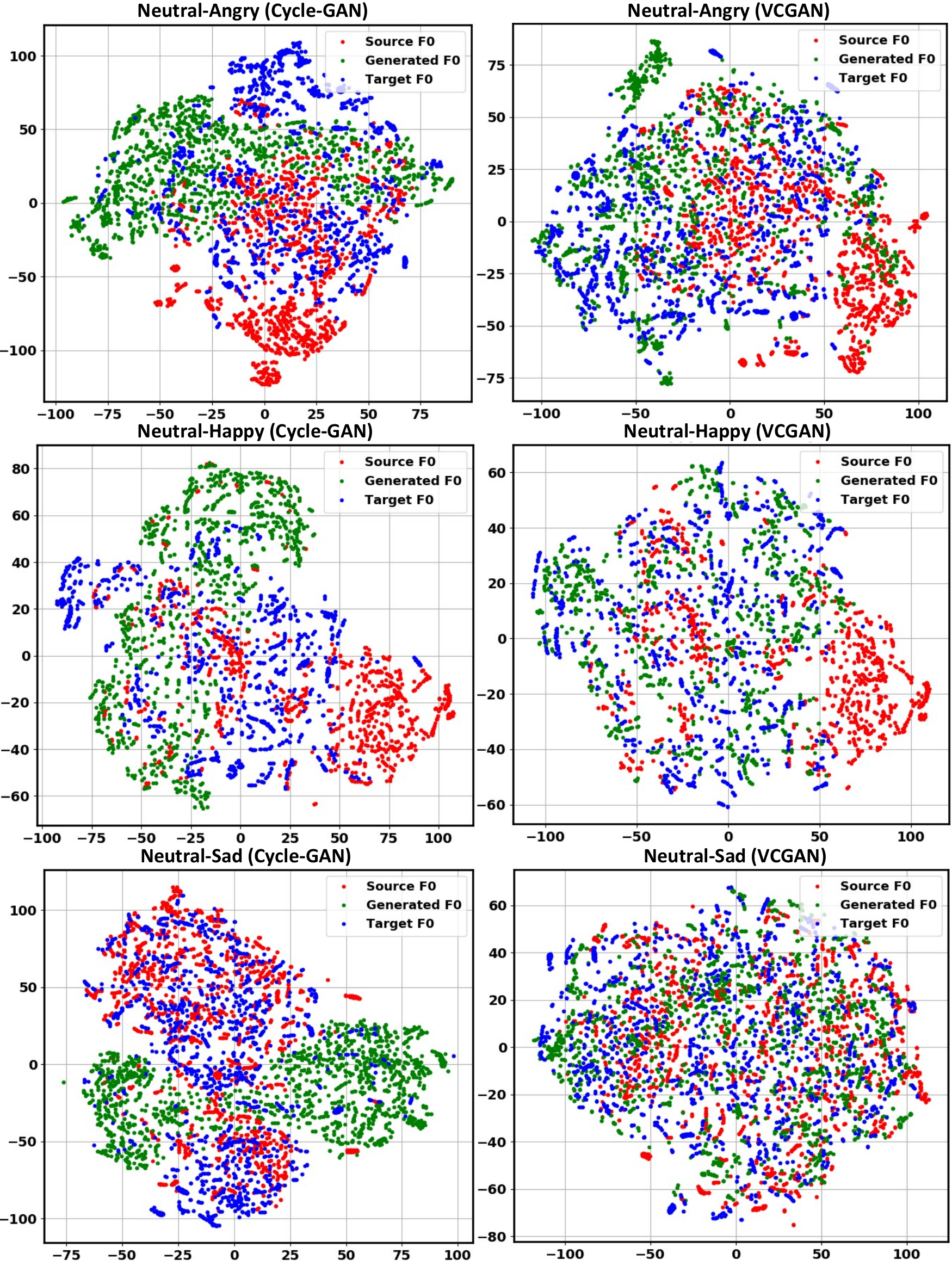}
  \caption{Visualizing t-SNE embeddings of source, converted and target F0 contours. The left column shows the embeddings generated using Cycle-GAN and the right column shows the same for variational model.}
  \label{fig:tsne_embeddings}
\vspace{-5mm}
\end{figure}

\subsection{Discriminator Loss and Architecture}
We model the probability ratio term in Eq.~(\ref{eqn:KL_upper_bound}) by a discriminator denoted by $D_{\gamma}$. Conceptually, this discriminator distinguishes between the joint distributions of $(\mathbf{S}_{A}, \mathbf{p}_{A})$ and $(\mathbf{S}_B, \mathbf{p}_{B})$ learned by generators $G_{\gamma}$ and $G_{\theta}$, respectively. 

During training of the discriminator $D_{\gamma}$, we minimize:
{\footnotesize
\begin{align} \nonumber
    & \mathscr{L}_{D_{\gamma}} = -E_{(\mathbf{S}_{A},\mathbf{p}_{A})} \Big[ E_{(\mathbf{S}_{B}, \mathbf{p}_{B}) \sim P_{\gamma}}  \Big[ \log \Big(  D_{\gamma}(\mathbf{S}_{A}, \mathbf{p}_{A}, \mathbf{S}_{B}, \mathbf{p}_{B}) \Big) \Big] \Big] \\ \label{eqn:discriminator_loss}
    & - E_{(\mathbf{S}_{B},\mathbf{p}_{B})} \Big[ E_{(\mathbf{S}_A, \mathbf{p}_{A}) \sim P_{\theta}}  \Big[ \log \Big(  1 - D_{\gamma}(\mathbf{S}_{A}, \mathbf{p}_{A}, \mathbf{S}_{B}, \mathbf{p}_{B}) \Big) \Big] \Big] 
\end{align}
}
A similar discriminator has been proposed to train autoencoders in an adversarial setting~\cite{veegan, adversarial_inference}. 
We use this discriminator to establish a macro connection between the two generators by providing them complete information about the generators. 
Another way to interpret Eq.~(\ref{eqn:discriminator_loss}) is that it classifies between different factorizations of the complete data distribution by the two generators. 
In fact, the optimal discriminators train the corresponding generators to minimize the Jensen-Shannon divergence between $P_{\gamma}(\mathbf{S}_{A}, \mathbf{p}_{A}, \mathbf{S}_{B}, \mathbf{p}_{B})$ and $P_{\theta}(\mathbf{S}_{A}, \mathbf{p}_{A}, \mathbf{S}_{B}, \mathbf{p}_{B})$ (derived in Appendix~B). 

We split each discriminator in two partial discriminators which separately provide the feedback for F0 and energy conversion task (see Appendix~B). There are three advantages to splitting up the discriminator's loss into an F0 and an energy contribution. First, this strategy provides greater flexibility, as the user can decide whether or not to perform energy conversion without altering the F0 transformation. This scenario may be useful in cases of limited training data, as we empirically observe greater variability in energy across utterances, thus making it harder to learn a conversion model. Second, as noted in this section, decoupling the F0 and energy backpropagation procedures prevents either variable from dominating the joint distribution during training. Third, we found empirically that training a unified discriminator results in an unstable model (see Appendix~C). This is because the gradient information must backpropagate through the energy generator to update the F0 model parameters. Thus, the error signal suffers from a vanishing gradient problem, which makes it challenging to properly train the generative models. In contrast, splitting the discriminator allows F0 model to get a direct feedback from its corresponding discriminator for improved learning.

We refer to this combined framework for F0 and energy conversion as a Variational Cycle-GAN (VCGAN).

\begin{algorithm*}
\setstretch{1.35}
\SetAlgoLined
\KwResult{Parameters $(\theta_G, \theta_D, \gamma_G,$ and $\gamma_D)$}
 \While{not converged}{
  \For{i = 1..n}{
      \textbf{Draw} $(\mathbf{S}_A^i, \mathbf{p}_A^i)$ and $(\mathbf{S}_B^i, \mathbf{p}_B^i)$ from emotion classes A and B\;
      \textbf{Compute} $\mathbf{e}_A^i$ and $\mathbf{e}_B^i$ from $\mathbf{S}_A^i$ and $\mathbf{S}_B^i$ \;
      \textbf{Sample} $\mathbf{m}_{p_{,AB}}^i$ and $\mathbf{m}_{p_{,BA}}^i$ using F0 momenta samplers \;
      \textbf{Generate} F0 contours $\mathbf{\Tilde{p}}_B^i$ and $\mathbf{\Tilde{p}}_A^i$ \;
      \textbf{Sample} $\mathbf{m}_{e_{,AB}}^i$ and $\mathbf{m}_{e_{,BA}}^i$ using energy momenta samplers \;
      \textbf{Generate} energy contours $\mathbf{\Tilde{e}}_B^i$ and $\mathbf{\Tilde{e}}_A^i$ and corresponding $\mathbf{\Tilde{S}}_B^i$ and $\mathbf{\Tilde{S}}_A^i$\;
      \textbf{Sample} $\mathbf{\hat{m}}_{p_{,AB}}^i$ and $\mathbf{\hat{m}}_{p_{,BA}}^i$ using the F0 momenta samplers for cyclic conversion\;
      \textbf{Generate} cycle converted F0 contours, $\mathbf{\hat{p}}_A^i$ and $\mathbf{\hat{p}}_B^i$\;
      \textbf{Sample} $\mathbf{\hat{m}}_{e_{,AB}}^i$ and $\mathbf{\hat{m}}_{e_{,BA}}^i$ using the energy momenta samplers for cyclic conversion\;
      \textbf{Generate} cycle converted energy contours $\mathbf{\Bar{e}}_A^i$ and $\mathbf{\Bar{e}}_B^i$\;
      \textbf{Sample} identity converted $\mathbf{\Bar{m}}_{e_{,AB}}^i$ and $\mathbf{\Bar{m}}_{e_{,BA}}^i$ using original spectrum and F0 contours\;
      \textbf{Generate} identity converted energy contours $\mathbf{\Bar{e}}_A^i$ and $\mathbf{\Bar{e}}_B^i$\;
  }
  $\nabla_{G_{\gamma}} \leftarrow \frac{1}{n} \sum_{i=1}^n [ \lambda_d \nabla\log(D_{\gamma}(\mathbf{S}_A^i, \mathbf{p}_A^i, \mathbf{\Tilde{S}}_B^i, \mathbf{\Tilde{p}}_B^i)) + \lambda_{c_1}\Arrowvert \mathbf{p}_A^i - \mathbf{\hat{p}}^i \Arrowvert_1 + \lambda_m(\Arrowvert \nabla \mathbf{m}_{AB}^i \Arrowvert^2 + \Arrowvert \nabla \mathbf{\hat{m}}_{AB}^i \Arrowvert^2 $
  $\phantom{\qquad \qquad \qquad \qquad \qquad \qquad \qquad} + \Arrowvert \nabla \mathbf{\Bar{m}}_{AB}^i \Arrowvert^2) + \lambda_i \Arrowvert \mathbf{e}_A^i - \mathbf{\Bar{e}}_A^i \Arrowvert_1 + \lambda_{c_2} \Arrowvert \mathbf{e}_A^i - \mathbf{\hat{e}}_A^i \Arrowvert_1]$ \;
  $\nabla_{G_{\theta}} \leftarrow \frac{1}{n} \sum_{i=1}^n [ \lambda_d \nabla\log(D_{\theta}(\mathbf{S}_B^i, \mathbf{p}_B^i, \mathbf{\Tilde{S}}_A^i, \mathbf{\Tilde{p}}_A^i)) + \lambda_{c_1}\Arrowvert \mathbf{p}_B^i - \mathbf{\hat{p}}_B^i \Arrowvert_1 + \lambda_m(\Arrowvert \nabla \mathbf{m}_{BA}^i \Arrowvert^2 + \Arrowvert \nabla \mathbf{\hat{m}}_{BA}^i \Arrowvert^2 $
  $\phantom{\qquad \qquad \qquad \qquad \qquad \qquad \qquad} + \Arrowvert \nabla \mathbf{\Bar{m}}_{BA}^i \Arrowvert^2) + \lambda_i \Arrowvert \mathbf{e}_B^i - \mathbf{\Bar{e}}_B^i \Arrowvert_1 + \lambda_{c_2} \Arrowvert \mathbf{e}_B^i - \mathbf{\hat{e}}_B^i \Arrowvert_1]$ \;
  $\nabla_{D_{\gamma}} \leftarrow \frac{1}{n} \sum_{i=1}^n [ -\nabla \log(D_{\gamma}(\mathbf{S}_A^i, \mathbf{p}_A^i, \mathbf{\Tilde{S}}_B^i, \mathbf{\Tilde{p}}_B^i)) -\nabla \log(1 - D_{\gamma}(\mathbf{\Tilde{S}}_A^i, \mathbf{\Tilde{p}}_A^i, \mathbf{S}_B^i, \mathbf{p}_B^i))]$\;
  $\nabla_{D_{\theta}} \leftarrow \frac{1}{n} \sum_{i=1}^n [ -\nabla \log(D_{\theta}(\mathbf{S}_B^i, \mathbf{p}_B^i, \mathbf{\Tilde{S}}_A^i, \mathbf{\Tilde{p}}_A^i)) -\nabla \log(1 - D_{\theta}(\mathbf{\Tilde{S}}_B^i, \mathbf{\Tilde{p}}_B^i, \mathbf{S}_A^i, \mathbf{p}_A^i))]$\;
 $\gamma_G \leftarrow \gamma_G - \eta_g \nabla_{G_{\gamma}}$\;
 $\theta_G \leftarrow \theta_G - \eta_g \nabla_{G_{\theta}}$\;
 $\gamma_D \leftarrow \gamma_D - \eta_d \nabla_{D_{\gamma}}$\;
 $\theta_D \leftarrow \theta_D - \eta_d \nabla_{D_{\theta}}$\;
 }
 \caption{Training procedure of VCGAN model}
\end{algorithm*}

\subsection{Modifying the Spectrum via Energy}

The spectral envelope is highly sensitive to changes in the location and filter response of the resonance frequencies. 
In fact, even minor changes can substantially degrade the quality and intelligibility of resynthesized speech. 
Our VCGAN framework circumvents this problem by modifying just the energy profile of the spectral envelope, i.e., the energy contour. 

First, we extract the energy contour of the given speech signal from its spectral representation using: 
\begin{equation}
\mathbf{e}_A = \sum_{f=0}^{\frac{F_s}{2}} [ \mathbf{S}_A ]_f^t
\end{equation}
where, $f$ corresponds to the frequency and $t$ is the time. 
Once the energy contour has been modified through the VCGAN, denoted as $\mathbf{e}_B$, then the converted spectrum $\mathbf{S}_B$ is given by:
\begin{equation}
\mathbf{S}_B = \mathbf{S}_A \times \frac{\mathbf{e}_B}{\mathbf{e}_A}
\end{equation}
This operation scales the frequency bins uniformly and simply modifies the overall intensity profile of the speech utterance. 

During training, we use 23-dimensional MFCC features for spectrum representation over a context of 128 frames extracted using a 5ms windows. 
The dimensionality of F0/energy contour is 128x1 while that of spectrum is 128x23. 
The smoothing kernel for registration is chosen to be $[6, 50]$ and $[6, 2]$ for the F0 and energy contour, respectively. 
The generator and discriminator networks are optimized alternately in every mini-batch update. 
We fix the mini-batch size to 2 and the learning rates are fixed at 1e-5 and 1e-7 for the generators and discriminators, respectively. 
We use Adam optimizer~\cite{adam_optimizer} with an exponential decay of 0.5 for the first moment. 
Sampling process in the generators is implemented via dropout~\cite{dropout} rate of 0.3 during both training and testing. 

\section{Experimental Results: Demonstrating Model Stability}

In this section, we demonstrate the desirable properties of our variational formulation, as compared to the traditional Cycle-GAN proposed in~\cite{cyclegan_emo_conv_2}. 
We also demonstrate the effectiveness of momenta regularization over the standard discrete wavelet transform representation. 
These experiments highlight the benefits of our VCGAN for emotion conversion.

\begin{figure*}[t]
  \centering
  \includegraphics[width=0.99\textwidth, height=5.0cm]{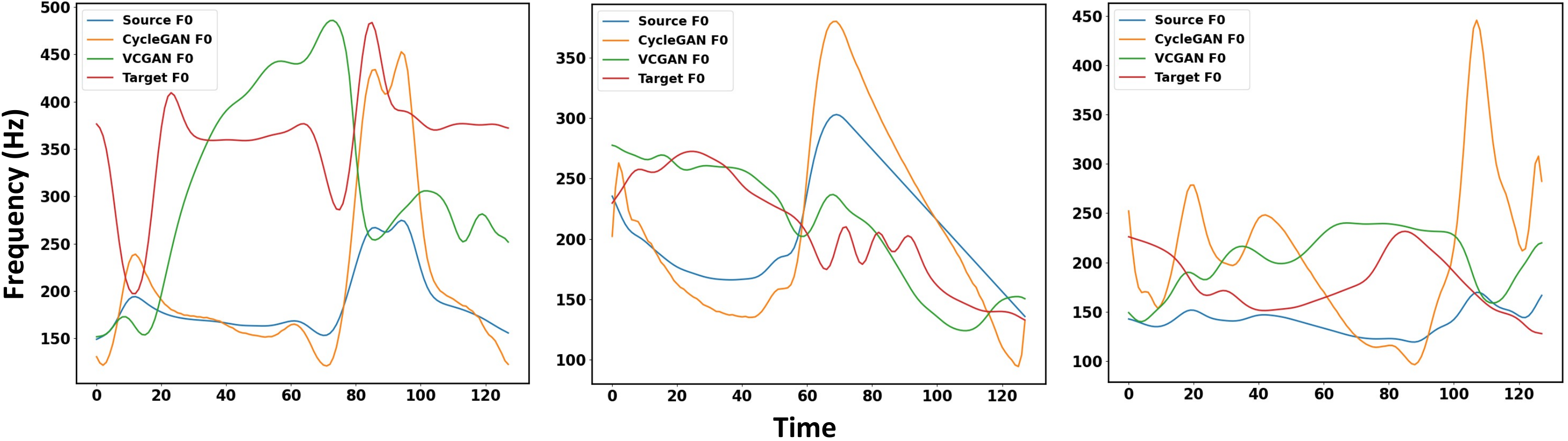}
  \caption{Comparing the F0 contours generated by Cycle-GAN and our momenta regularized variational model. Using diffeomorphic warping as a regularizer leads to more stable F0 contour generation in comparison to wavelet based regularization.}
  \label{fig:comparing_f0s}
\end{figure*}

\begin{figure}[!t]
  \centering
  \includegraphics[width=0.88\linewidth, height=5.2cm]{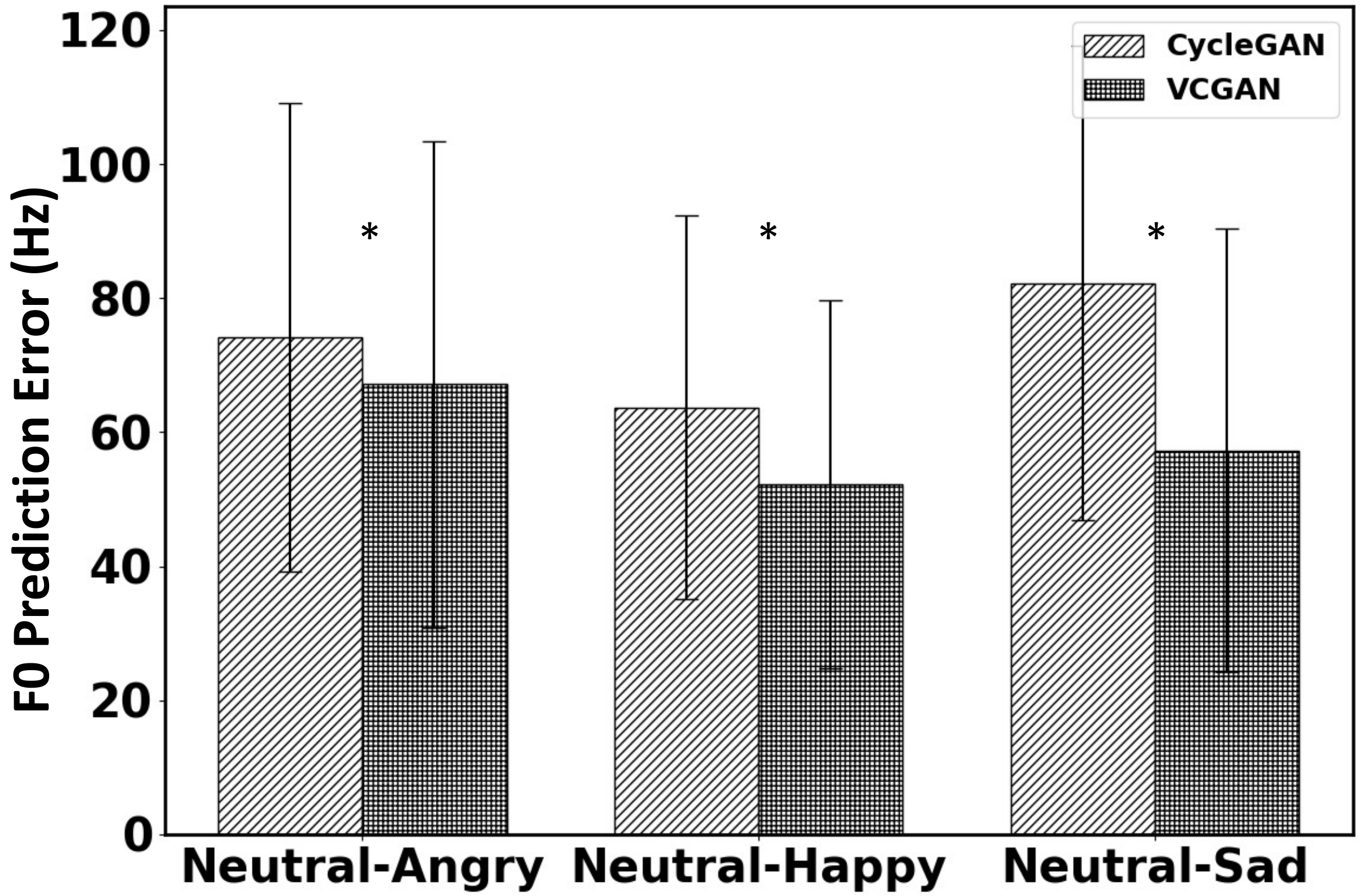}
  \caption{F0 RMSE comparison between Cycle-GAN and VCGAN. The results are statistically significant at level 0.05 ($\text{*}$ denote p-value $\leq 1e-10$).}
  \label{fig:pitch_rmse_barplot}
\vspace{-2mm}
\end{figure}

\begin{figure}[!t]
  \centering
  \includegraphics[width=0.88\linewidth, height=5.2cm]{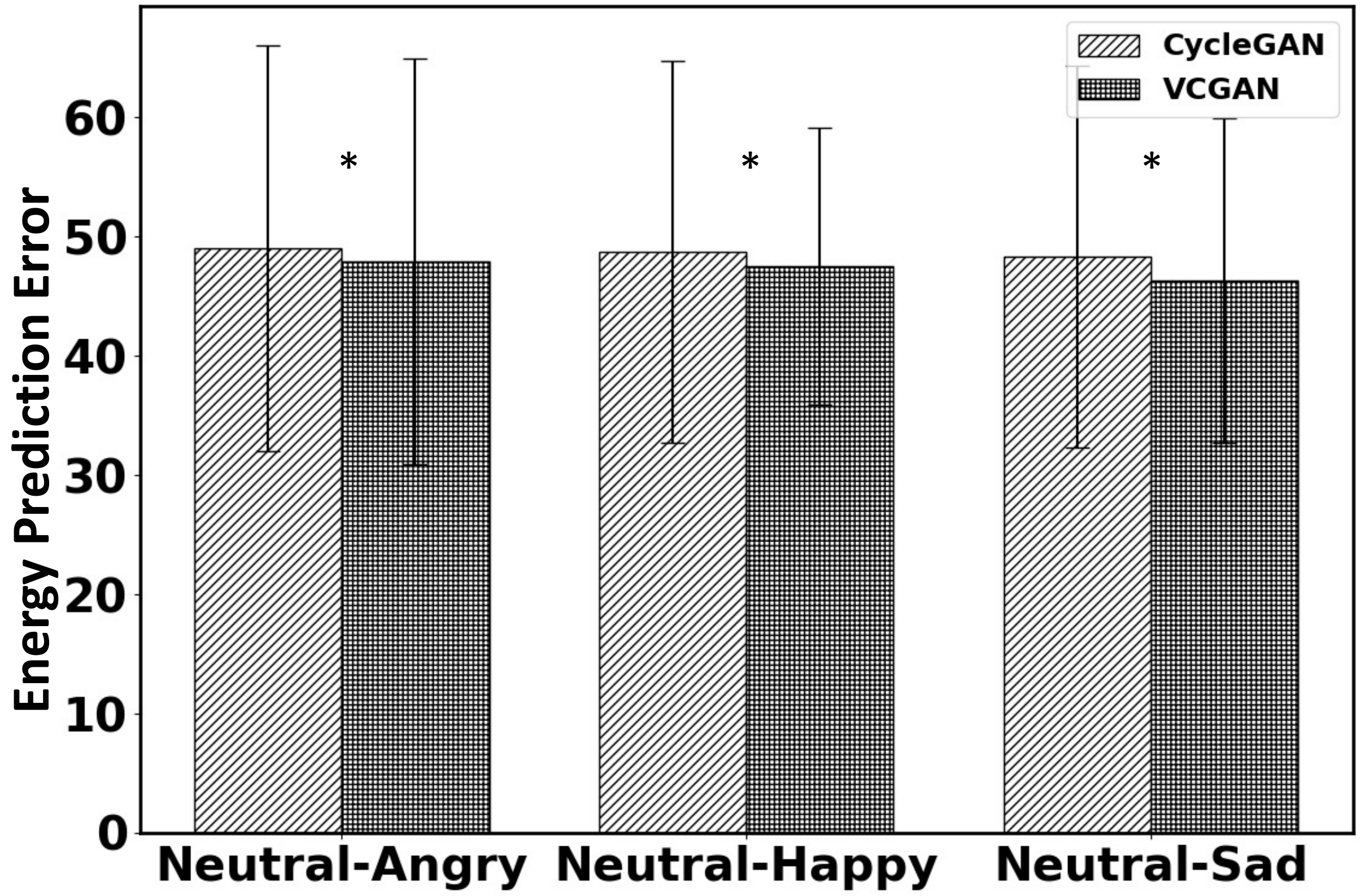}
  \caption{Energy RMSE comparison between Cycle-GAN and VCGAN. Results are statistically significant at level 0.05 ($\text{*}$ denote p-value $\leq 1e-10$).}
  \label{fig:energy_rmse_barplot}
\vspace{-5mm}
\end{figure}

\subsection{VESUS Dataset}
We evaluate our algorithms on the VESUS dataset~\cite{vesus} collected at Johns Hopkins University. 
VESUS contains 250 utterances/phrases spoken by 10 different actors (gender balanced) in neutral, sad, angry and happy emotional classes. 
Each spoken utterance has a crowd-sourced emotional saliency rating collected from 10 workers on Amazon Mechanical Turk (AMT)~\cite{amt}. 
These ratings represent the ratio of workers who correctly identify the intended emotion in a recorded utterance. 
For robustness, we restrict our experiments in this section and the next to utterances that were correctly and consistently rated as emotional by at least 5 out of the 10 AMT workers. 
The total number of utterances for each emotion class are:
\begin{itemize}
    \item \textbf{Neutral to Angry conversion}: 1667 utterances.
    \item \textbf{Neutral to Happy conversion}: 876 utterances.
    \item \textbf{Neutral to Sad conversion}: 1587 utterances.
\end{itemize}

\begin{figure*}[t]
  \centering
  \includegraphics[width=0.93\textwidth, height=9.5cm]{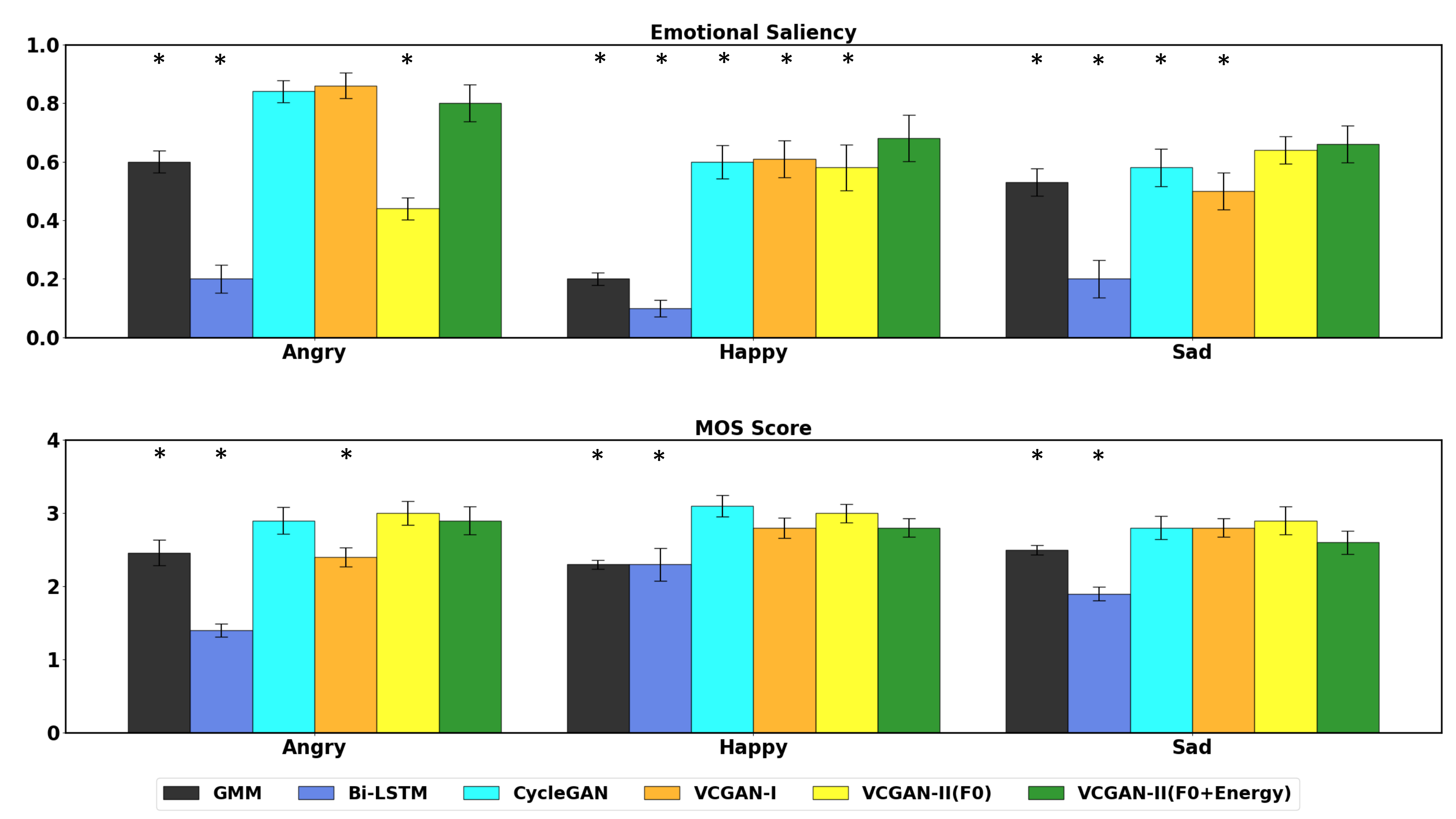}
  \caption{\textbf{Single-Speaker Evaluation:} we create the training, validation and testing sets for each emotion pair based on the speaker from VESUS with the highest number of emotionally salient utterances. The asterisk ($^*$) denotes statistical significance for the test (VCGAN-II (F0+Energy) $>$ Method) at $p<0.05$.}
  \label{fig:vesus_spk5}
\end{figure*}

\subsection{Stability of Training}
We first evaluate model stability during training. 
Here, we borrow from game theory to quantify performance. 
Namely, the optimal outcome of an adversarial game occurs when both participants achieve the Nash equilibrium~\cite{nash_equilibrium}. 
Translating this idea into generative adversarial training implies equality of generator and discriminator losses. 
While a strict equality is difficult to achieve in practice, similar losses typically indicate better quality of the generated samples. 
Fig.~\ref{fig:stability_result} shows the difference between the generator and discriminator losses for the Cycle-GAN (orange) and our proposed VCGAN (blue). 
We note that the VCGAN achieves better calibration of the generator and discriminator objectives (i.e., near equality), whereas traditional Cycle-GAN fails to do so. 
Thus, we conclude that our training algorithm exhibits better stability in practice. 
Another important aspect of our proposed strategy is that computed training loss wiggles around the optimal point. 
It is crucial for adversarial training as the absence of this variance can sometimes signal mode collapse~\cite{mode_collapse}. 

To illustrate the improved generator calibration, Fig.~\ref{fig:tsne_embeddings} shows the tSNE plots~\cite{tsne} of the source, generated, and target emotion F0 values extracted over 640ms long windows. This duration typically encompasses multiple syllables in conversational English, often corresponding to words, and is therefore supra-segmental in nature. 
Notice that the point cloud of generated F0 values by the Cycle-GAN shows poor overlap with the target F0 distribution. 
We hypothesize that, as the Cycle-GAN focuses on just the first-order moments, the generators ultimately learn a mapping function whose output lies on a completely different manifold than the actual data distribution. 
This further indicates that the Cycle-GAN acts as a poor estimator of the target data density due to the weak constraint imposed by cycle-consistency loss. 
The VCGAN, on the other hand, does a much better job of approximating the target data density. 
This is because the KL-divergence penalty between the given data distribution and its cyclic counterpart enforces a stronger global dependency between the two generators. 
This macro connection in the form of feedback from the joint-density discriminator facilitates learning a better mapping function, especially given the limited data.

\begin{figure*}[t]
  \centering
  \includegraphics[width=0.93\textwidth, height=9.5cm]{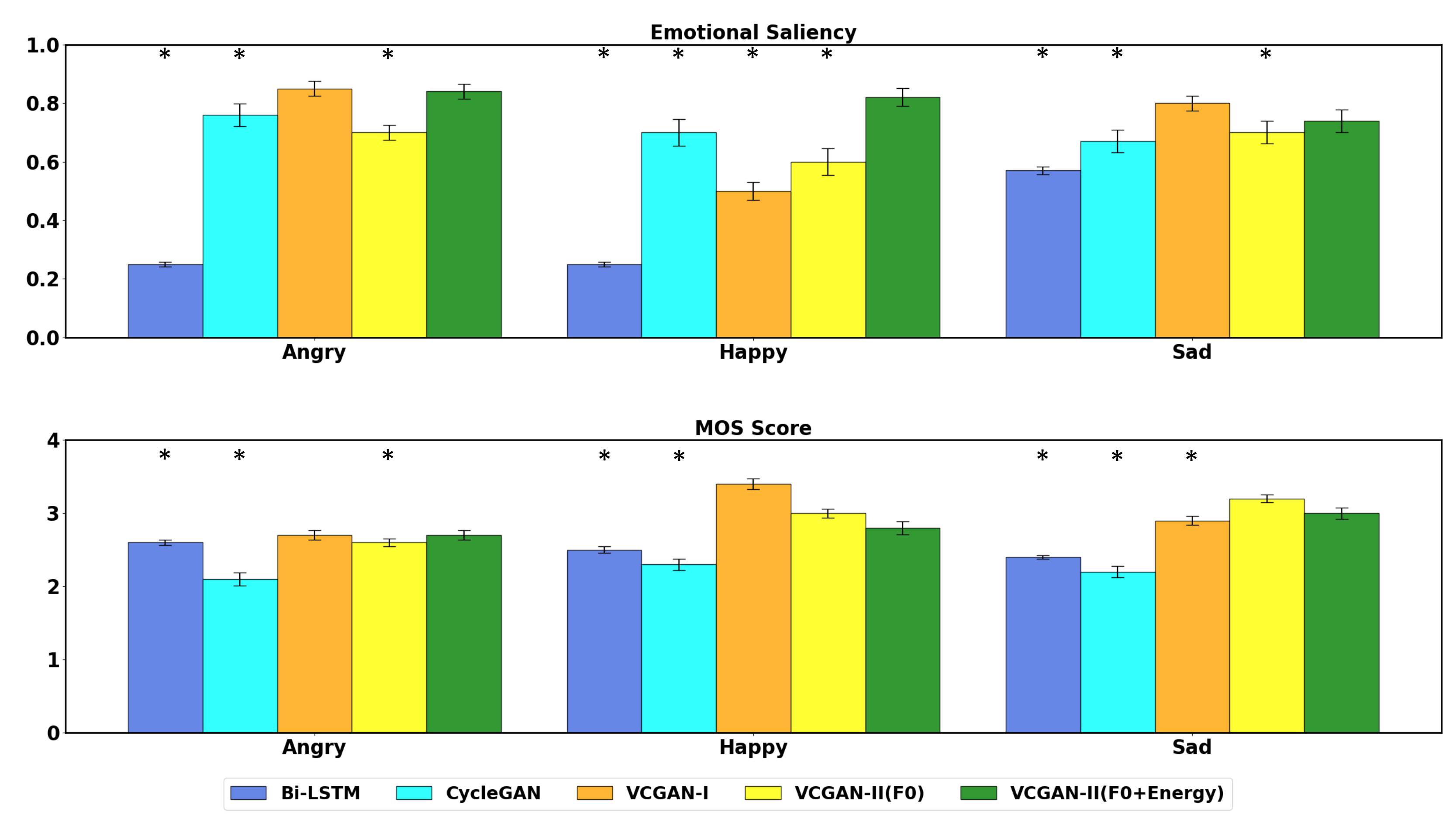}
  \caption{\textbf{Mixed Speaker Evaluation:} we create the training, validation and testing sets by randomly sampling utterances from VESUS across all speakers. The asterisk ($^*$) denotes statistical significance for the one-tailed t-test (VCGAN (F0+Energy) $>$ Method) at $p<0.05$.}
  \label{fig:vesus_mixed}
\end{figure*}

\subsection{Effect of Momenta Regularization}
The second critical component of our proposed VCGAN framework is the momenta based regularization for modeling the target prosodic contours. 
As discussed, the momenta specify an iterative warping process. 
In contrast, the works of~\cite{lstm_emo_conv, cyclegan_emo_conv} use a continuous wavelet transform to parameterize the F0 contour to stabilize its generative process. 
Empirically, we observe that our proposed momenta-based warping allows more flexible transformations and better scale matching between the generated and target contours. 
Fig.~\ref{fig:comparing_f0s} shows example pitch contours \textit{generated during testing}. 
As seen, the momenta-based VCGAN is less sensitive to extreme local fluctuations in the generated contours due to the iterative warping process. 
Moreover, our warping approach takes the source F0 contour as a baseline curve and estimates a perturbation on top of it. 
This results in a better alignment of the scale of F0 values in going from one emotion to the other (see Fig.~\ref{fig:comparing_f0s}). 

To establish our claim objectively, we use paired samples from the VESUS dataset to compute root mean square error (RMSE) between the generated and target frames of the F0 (Fig.~\ref{fig:pitch_rmse_barplot}) and energy (Fig.~\ref{fig:energy_rmse_barplot}) contours. 
As seen, our momenta-based warping is significantly better than wavelet based regularization used in~\cite{lstm_emo_conv, cyclegan_emo_conv_2} for all three emotion conversion tasks. 
The overall F0 loss is slightly higher for neutral-angry and neutral-sad conversion in comparison to the neutral-happy conversion. 
This is because the sad and angry emotions are portrayed in a more diverse manner in the VESUS dataset. 

\section{Experimental Results: Emotion Conversion}

In this section we evaluate the emotion conversion performance against several supervised and unsupervised baseline algorithms. 
We train a separate model for each pair of emotions. However, the model architecture remains fixed in each case.
Our subjective evaluation includes both an emotion perception query and a quality assessment test carried out on Amazon Mechanical Turk (AMT). 
Specifically, each pair of speech utterances (neutral and converted) is rated by 5 workers on AMT. 
The perception test asks the raters to identify the emotion in the converted speech sample after listening the corresponding neutral utterance. 
The quality assessment test asks them to rate the quality of the speech sample on a 1-5 scale, also called as mean opinion score or MOS. 
The reason we include both the neutral and converted utterances is to account for the speaker bias. 
Given the known variability in emotional perception across people, we collect $5$ ratings for each converted sample and report the average.
Finally, some samples were randomly and intentionally corrupted to mitigate the effects of non-diligent raters and to identify/flag bots. 

We conduct four evaluations of increasing level of difficulty. The simplest scenario is single-speaker emotion conversion, in which we train and evaluate the model on utterances from the same speaker. Next is a mixed-speaker evaluation, in which we pool the utterances across speakers for each emotion class and randomly divide them into training, validation, and testing. The third assessment is out-of-speaker evaluation; here the models are trained and tested on different speakers. Finally, our Wavenet evaluation is the most difficult and queries how well the models generalize to synthetic speech.

\subsection{Baseline Models}
We compare our proposed VCGAN with several state-of-the-art algorithms from supervised and unsupervised learning domains. 
The first baseline is the global variance constrained GMM used for voice conversion, which learns the join density of source and target emotion features~\cite{gmm_emo_conv}. 
The second baseline uses a Bi-LSTM model~\cite{lstm_emo_conv} to learn the conditional density of the target emotion features namely, the F0 and energy contours. 
This method uses the wavelet decomposition of the prosodic features to control the segmental and supra-segmental nature of prosody. 
The third technique is a recently proposed Cycle-GAN framework~\cite{cyclegan_emo_conv} to modify the F0 contour using its wavelet parameterization. 
Further, the authors learn a secondary set of Cycle-GANs to modify the mel-cepstral features for every pair of source-target emotions. 
Our fourth baseline is a simplified version of the proposed VCGAN model~\cite{vcgan} (referred in experiments as VCGAN-I). 
It is a mixed approach in the sense that it learns a variational Cycle-GAN for the F0 conversion and a traditional Cycle-GAN for converting the Mel-cepstral features. 
In essence, all of the baseline techniques in this work modify the F0 and energy contour (as extracted from the spectrum or MFCC features).
Finally, we compare our complete F0+energy modification framework with just F0 modification to understand the role of energy contour. We refer the reader to the Appendix~C for detail descriptions of each baseline architecture, including size and number of parameters.

\subsection{Single Speaker Evaluation}
We first evaluate how well our VCGAN framework can convert emotions for a single speaker. Note that, this is the simplest setting in which our goal is to show generalization on a single speaker. To maximize the amount of data, we select the VESUS speaker with the highest number of consistently rated utterances (see Section~III-A) for each emotion pair. This yields the following sample sizes:
\begin{itemize}
    \item \textbf{Neutral to Angry Conversion}: 200 utterances for training, 25 for validation and, 10 for testing.
    \item \textbf{Neutral to Happy Conversion}: 100 utterances for training, 5 for validation and, 10 for testing.
    \item \textbf{Neutral to Sad Conversion}: 200 utterances for training, 25 for validation and, 10 for testing.
\end{itemize}

Fig.~\ref{fig:vesus_spk5} illustrates the performance across all models in this single speaker setting. We notice that the Bi-LSTM suffers due to the limited training utterances, which suggests that the model cannot learn an appropriate mapping with this amount of data.
GMM model fares better because it has the least amount of parameters among all the competing methods. 
It is capable of learning some aspects of the transfer function in a data-starved scenario.
The Cycle-GAN achieves comparable performance to VCGAN-II (F0+Energy) on emotional saliency and outperforms our method on the MOS score. 
This behavior is unsurprising, as the Cycle-GAN architecture was designed for and evaluated on single speaker conversion tasks.
VCGAN-II(F0+Energy) achieves the most robust performance across the three VCGAN models. We posit that this may be due to its reduced parameterization and focus on both F0 and energy.

\begin{figure*}[t]
  \centering
  \includegraphics[width=0.93\textwidth, height=9.5cm]{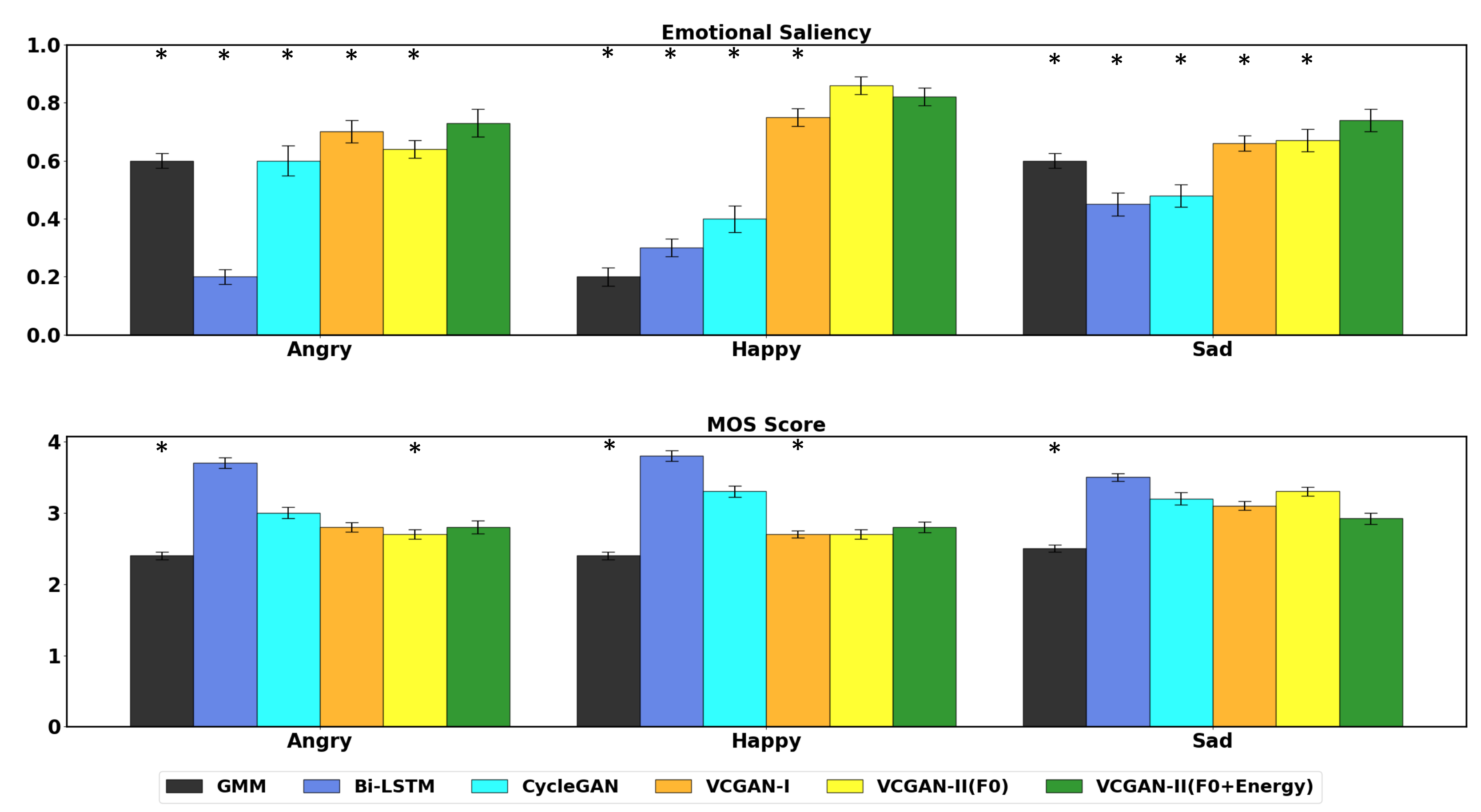}
  \caption{\textbf{Out-of-Speaker Evaluation:} we create 5 folds from VESUS, each comprising of a male and a female speaker. We train the VCGAN model on four folds and evaluate its performance on the fifth. The asterisk ($^*$) denotes statistical significance for the test (VCGAN-II (F0+Energy) $>$ Method) at $p<0.05$.}
  \label{fig:vesus_folds}
\end{figure*}

\begin{figure*}[t]
  \centering
  \includegraphics[width=0.93\textwidth, height=9.5cm]{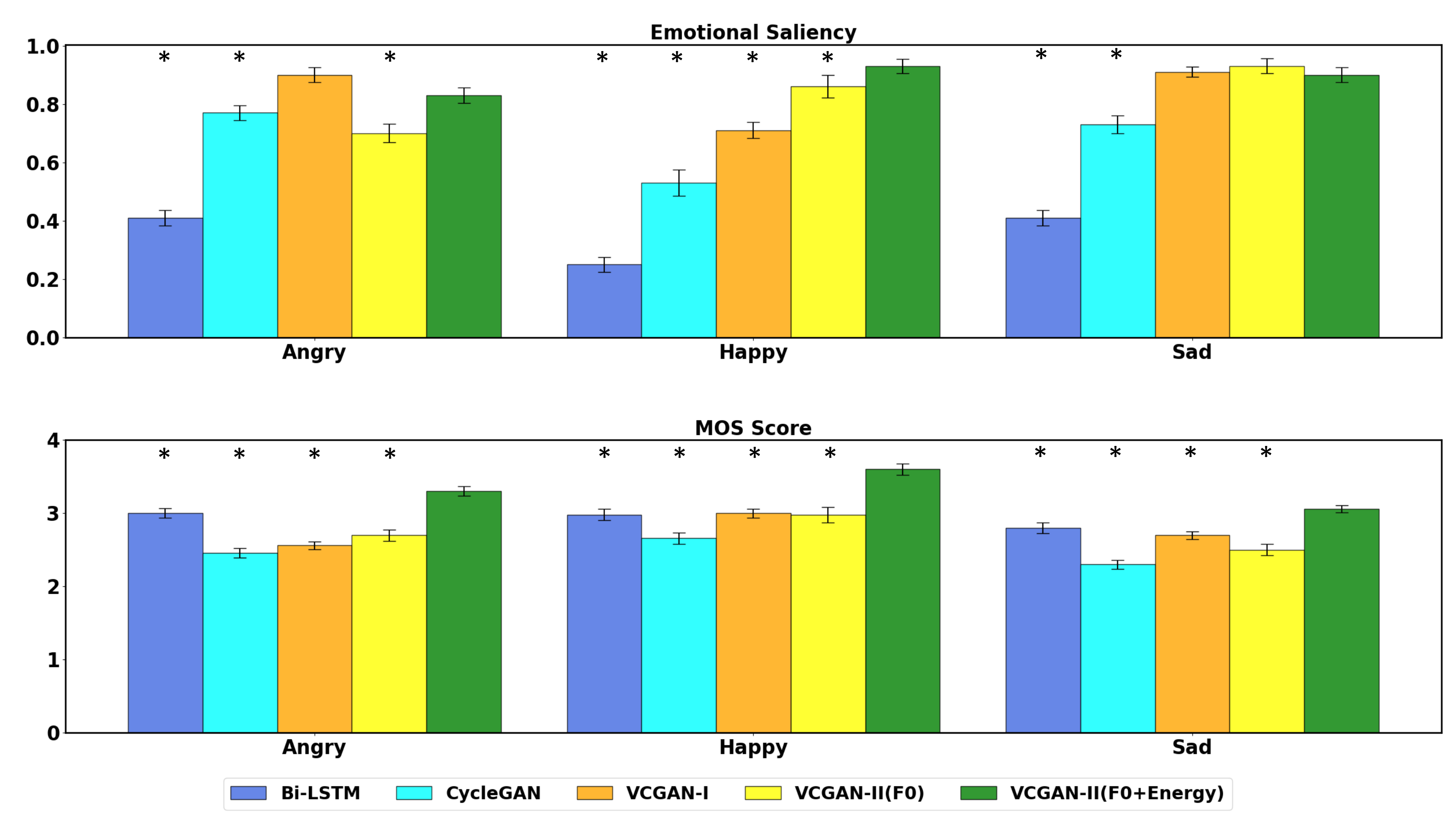}
  \caption{\textbf{Wavenet Evaluation:} We apply our mixed speaker models (without fine-tuning) to modify speech generated by the Wavenet model. The asterisk ($^*$) denotes statistical significance for the one-tailed t-test (VCGAN (F0+Energy) $>$ Method) at $p<0.05$.}
  \label{fig:wavenet_mixed}
\end{figure*}

\subsection{Mixed Speaker Evaluation}
To evaluate the performance of our model in a mixed speaker setting, we split the VESUS corpus as follows:
\begin{itemize}
    \item \textbf{Neutral to Angry Conversion}: 1534 utterances for training, 72 for validation and, 61 for testing.
    \item \textbf{Neutral to Happy Conversion}: 790 utterances for training, 43 for validation and, 43 for testing.
    \item \textbf{Neutral to Sad Conversion}: 1449 utterances for training, 75 for validation and, 63 for testing.
\end{itemize}
We use the training and validation set to learn the parameters of the models and then evaluate using the test set. 
Empirically, we observe that the GMM does not produce intelligible speech in this setting due to the wide variation of speakers. 
Therefore, we have removed from the analysis. Fig.~\ref{fig:vesus_mixed} shows the results of the crowd-sourcing experiment on test dataset. 
The mixed speaker setting is more challenging than the single speaker case because of variability across speakers in terms of estimating the dynamic range of the prosodic features. 
There is a high chance of learning an average mapping by the model. 

We note that the Bi-LSTM has the worst emotion conversion accuracy (below 50$\%$) for all three pairs of emotions. However, it also generates reasonably good audio quality. 
Empirically, we observe that the Bi-LSTM learns an near-identity mapping, meaning it does not perform any emotion conversion, but simply reconstructs the (already high quality) input utterance. 
The Cycle-GAN~\cite{cyclegan_emo_conv_2} model fairs reasonably well in terms of emotional saliency; however, the speech reconstruction quality is significantly lower than all three of our proposed models. 
VCGAN-II (F0 and F0+energy), in comparison, shows a uniformly consistent performance across the three emotion classes and does an extremely good job of retaining the speech naturalness post-conversion. 
We attribute this all-round performance to the momenta based regularization and to the variational formulation. 
The VCGAN-I model comes close to our proposed F0+energy framework for angry and sad emotions, but its conversion accuracy falls below 50$\%$ for the happy emotion, making it the least consistent.

\subsection{Out-of-Speaker Evaluation}
We now tackle the more challenging task of out-of-speaker generalization. 
Here, we create five folds from VESUS, each one consisting of a single male and a single female speaker. We then train five separate models for each neutral $\Longleftrightarrow$ emotional pair corresponding using four of these folds and then test on the fifth remaining fold. 
Note that, this task tests the model's ability to generalize and learn transformation for speakers which are not part of the training set. Since each speaker has a different number of consistently rated emotional utterances, the data splits are fold-dependent, as shown in Table~\ref{tab:fold_eval}.

\begin{table}[t]
    \centering
    \caption{Data splits used for the Out-of-Speaker Evaluation}
    \label{tab:fold_eval}
    \begin{tabular}{ |c|c|c|c|c| } 
     \hline
     \textbf{Emotion Pair} & \textbf{Fold} & \textbf{Train} & \textbf{Validation} & \textbf{Test} \\ 
     \hline
     \multirow{5}{6em}{Neutral-Angry} & 1 & 1347 & 320 & 123 \\
     & 2 & 1212 & 455 & 125 \\
     & 3 & 1610 & 57 & 122 \\
     & 4 & 1310 & 357 & 125 \\ 
     & 5 & 1189 & 478 & 107 \\
     \hline
     \multirow{5}{6em}{Neutral-Happy} & 1 & 710 & 166 & 285 \\
     & 2 & 581 & 295 & 289 \\
     & 3 & 779 & 97 & 278 \\
     & 4 & 833 & 43 & 284 \\ 
     & 5 & 601 & 275 & 220 \\
     \hline
     \multirow{5}{6em}{Neutral-Sad} & 1 & 1357 & 230 & 169 \\
     & 2 & 1340 & 247 & 174 \\
     & 3 & 1154 & 433 & 172 \\
     & 4 & 1329 & 258 & 173 \\ 
     & 5 & 1167 & 420 & 153 \\
     \hline
    \end{tabular}
\end{table}

We sample 10 utterances for each fold and each emotion pair to collect the final ratings. 
Fig.~\ref{fig:vesus_folds} shows the average performance across the folds for all methods. 
Once again, we evaluate two variants of our proposed framework: VCGAN-II(F0) and VCGAN-II(F0+Energy). 
Once again, the GMM fails to produce intelligible speech for the out-of-speaker experiment. 
Therefore, we have trained it for each speaker individually rather than fold-wise. 
Ultimately, the GMM model is not suitable for a real-world application, where the speakers may be unknown or vary between training and deployment.

At a first glance, we can see that the unsupervised models (GANs) generally outperforms the supervised method (Bi-LSTM). 
In fact, the Bi-LSTM model has the worst emotion conversion accuracy (below 50$\%$) for all three pairs of emotions. 
However, it also generates the best audio quality among all the competing models, likely due to the minimal conversion. 
This result suggests a trade-off, which requires balancing the ``strength" of the emotion conversion but not distorting the spectrum and F0 contour by too much modification.

Among the unsupervised models, Cycle-GAN has uniformly poor conversion accuracy across all emotion pairs. 
It fails to generalize to unseen speakers due to its weak generator-discriminator coupling and the variation between training and testing speakers. 
The generated speech quality is however higher, which is consistent with the behavior of the Bi-LSTM model. 
VCGAN models (VCGAN-I and VCGAN-II) outperform the remaining models in the fold-wise evaluation in emotion conversion task. 
They also achieve a good trade-off between emotion conversion and maintaining the naturalness of speech. 
The VCGAN-II(F0+Energy) model has the best balance among the three and is consistent on 2 out of 3 emotion conversion tasks. 
The difference between VCGAN-II(F0) and VCGAN-II(F0+Energy) demonstrates how variation in energy plays an important role in the perception of emotion. 
Angry and sad emotions seem to be affected the most by this variation. 
Angry emotion is often characterized by a significant rise in the loudness, whereas sad emotion is exactly the opposite. 
Our VCGAN-II(F0+Energy) model is able to capture and encapsulate this information to some extent.  

\subsection{Wavenet Evaluation}
Our final evaluation is on synthetic speech. In this case, we use the models trained in the mixed speaker evaluation (Section~IV-C) without any fine tuning. This paradigm is more challenging because the test speaker characteristics are completely different from the training set. We generate ``neutral" utterances using the Wavenet API provided by Google~[43]. The utterances are based on randomly sampled phrases from the VESUS dataset to preserve syntactic similarity between training and testing. The number of testing utterances is the same as in Section~IV-C: 61 for neutral $\rightarrow$ angry, 43 for neutral $\rightarrow$ happy, and 63 for neutral $\rightarrow$ sad. Since, the Wavenet model generates audio in time domain directly, we use the WORLD vocoder to extract acoustic and prosodic features.

Fig.~\ref{fig:wavenet_mixed} shows that our VCGAN-II models are extremely good at converting the emotions in synthesized speech. 
While VCGAN-I matches the proposed model in terms of emotional saliency, the quality of generated audio trends significantly lower in comparison. 
This is likely due to the secondary spectrum modification, which is not harmonically matched with the modified F0 contours. 
This experiment further demonstrates our VCGAN-II framework is robust even when the unseen speaker has completely different characteristics than the dataset on which the model has been trained. 
This is a first model in our knowledge that generalizes so well to a synthetic speaker (simulated by Wavenet). 

\subsection{Summary of Results}
Table~\ref{tab:three_paradigms} summarizes the crowd sourcing results across the different evaluation paradigms, i.e., single speaker, mixed speaker, out-of-speaker, and Wavenet. 
Right away, we observe an inconsistency in performance as we progress from one experiment setting to another. This variation is expected, due to the increasing levels of difficulty of each evaluation. Specifically, our single speaker evaluation queries the performance of each model on utterances from the same speaker. In the mixed-speaker case, we train and test the models on the same collection of speakers, but randomly split the utterances between the two sets. This evaluation is more challenging because the models must learn characteristics of multiple speakers. In the out-of-speaker evaluation, we train the models on a set of four male/female speaker pairs and test on the remaining pair. Thus, the models never see utterances from the test speakers during training, which is a more difficult task. Finally, the Wavenet evaluation queries how well the models generalize to synthetic speech, which by default is produced under different environmental (and physiological) conditions.

\begin{table*}[t]
\begin{center}
\caption{Performance across the \textcolor{blue}{four} evaluation paradigms: Single-speaker, Mixed-speaker, Out-of-speaker, and Wavenet.}
\begin{tabular}{ |c|c|c|c|c|c|c|c| }
\hline
Evaluation & Algorithm & \multicolumn{2}{c|}{Neutral-angry} & \multicolumn{2}{c|}{Neutral-happy} & \multicolumn{2}{c|}{Neutral-sad} \\
& & Acc. & MOS & Acc. & MOS & Acc. & MOS \\
\hline
\multirow{6}{4em}{Single Speaker} & GMM~\cite{gmm_emo_conv} & 0.6$\pm$0.1 & 2.5$\pm$0.6 & 0.2$\pm$0.1 & 2.3$\pm$0.2 & 0.53$\pm$0.2 & 2.5$\pm$0.4\\
& Bi-LSTM~\cite{lstm_emo_conv} & 0.2$\pm$0.2 & 1.4$\pm$0.3 & 0.1$\pm$0.1 & 2.3$\pm$0.7 & 0.2$\pm$0.2 & 1.9$\pm$0.3\\
& Cycle-GAN~\cite{cyclegan_emo_conv_2} & 0.84$\pm$0.1 & 2.9$\pm$0.6 & 0.6$\pm$0.2 & 3.1$\pm$0.5 & 0.58$\pm$0.2 & 2.8$\pm$0.5\\
& VCGAN-I~\cite{vcgan} & \textbf{0.86$\pm$0.1} & 2.4$\pm$0.4 & 0.6$\pm$0.2 & 2.8$\pm$0.4 & 0.5$\pm$0.2 & 2.8$\pm$0.4\\ 
& VCGAN-II(F0) & 0.44$\pm$0.1 & \textbf{3.0$\pm$0.5} & 0.58$\pm$0.2 & \textbf{3.0$\pm$0.4} & 0.64$\pm$0.2 & \textbf{2.9$\pm$0.6}\\
& VCGAN-II(F0+Energy) & 0.8$\pm$0.2 & 2.9$\pm$0.6 & \textbf{0.68$\pm$0.2} & 2.8$\pm$0.4 & \textbf{0.66$\pm$0.2} & 2.6$\pm$0.5\\
\hline
\multirow{5}{3em}{Mixed Speaker} & Bi-LSTM~\cite{lstm_emo_conv} & 0.25$\pm$0.1 & 2.6$\pm$0.3 & 0.25$\pm$0.1 & 2.5$\pm$0.3 & 0.57$\pm$0.1 & 2.4$\pm$0.2\\
& Cycle-GAN~\cite{cyclegan_emo_conv_2} & 0.76$\pm$0.3 & 2.1$\pm$0.7 & 0.7$\pm$0.3 & 2.3$\pm$0.5 & 0.67$\pm$0.3 & 2.2$\pm$0.6\\
& VCGAN-I~\cite{vcgan} & \textbf{0.85$\pm$0.2} & \textbf{2.7$\pm$0.5} & 0.5$\pm$0.2 & \textbf{3.4$\pm$0.5} & 0.8$\pm$0.2 & 2.9$\pm$0.5\\ 
& VCGAN-II(F0) & 0.7$\pm$0.2 & 2.6$\pm$0.4 & 0.6$\pm$0.3 & 3.0$\pm$0.4 & 0.7$\pm$0.3 & \textbf{3.2$\pm$0.4}\\
& VCGAN-II(F0+Energy) & 0.84$\pm$0.2 & \textbf{2.7$\pm$0.5} & \textbf{0.82$\pm$0.2} & 2.8$\pm$0.6 & \textbf{0.74$\pm$0.3} & 3.0$\pm$0.6\\
\hline
\multirow{6}{4em}{Out-of-Speaker} & GMM~\cite{gmm_emo_conv} & 0.6$\pm$0.2 & 2.4$\pm$0.4 & 0.2$\pm$0.2 & 2.4$\pm$0.4 & 0.6$\pm$0.2 & 2.5$\pm$0.4\\
& Bi-LSTM~\cite{lstm_emo_conv} & 0.2$\pm$0.2 & \textbf{3.7$\pm$0.6} & 0.3$\pm$0.2 & \textbf{3.8$\pm$0.6} & 0.45$\pm$0.3 & \textbf{3.5$\pm$0.4}\\
& Cycle-GAN~\cite{cyclegan_emo_conv_2} & 0.6$\pm$0.4 & 3.0$\pm$0.6 & 0.4$\pm$0.3 & 3.3$\pm$0.6 & 0.48$\pm$0.3 & 3.2$\pm$0.7\\
& VCGAN-I~\cite{vcgan} & 0.7$\pm$0.3 & 2.8$\pm$0.5 & 0.75$\pm$0.2 & 2.7$\pm$0.4 & 0.66$\pm$0.2 & 3.1$\pm$0.5\\ 
& VCGAN-II(F0) & 0.64$\pm$0.2 & 2.7$\pm$0.5 & \textbf{0.86$\pm$0.2} & 2.7$\pm$0.5 & 0.67$\pm$0.3 & 3.3$\pm$0.5\\
& VCGAN-II(F0+Energy) & \textbf{0.73$\pm$0.3} & 2.8$\pm$0.7 & 0.82$\pm$0.2 & 2.8$\pm$0.6 & \textbf{0.74$\pm$0.3} & 2.9$\pm$0.6\\
\hline
\multirow{5}{3em}{Wavenet} & Bi-LSTM~\cite{lstm_emo_conv} & 0.41$\pm$0.2 & 3.0$\pm$0.5 & 0.25$\pm$0.2 & 2.98$\pm$0.5 & 0.41$\pm$0.2 & 2.8$\pm$0.6\\
& Cycle-GAN~\cite{cyclegan_emo_conv_2} & 0.77$\pm$0.2 & 2.46$\pm$0.5 & 0.53$\pm$0.3 & 2.66$\pm$0.5 & 0.73$\pm$0.2 & 2.3$\pm$0.5\\
& VCGAN-I~\cite{vcgan} & \textbf{0.9$\pm$0.2} & 2.56$\pm$0.4 & 0.71$\pm$0.2 & 3.0$\pm$0.4 & 0.9$\pm$0.1 & 2.7$\pm$0.4\\ 
& VCGAN-II(F0) & 0.7$\pm$0.24 & 2.7$\pm$0.6 & 0.86$\pm$0.3 & 2.98$\pm$0.7 & \textbf{0.93$\pm$0.2} & 2.5$\pm$0.6\\
& VCGAN-II(F0+Energy) & 0.83$\pm$0.2 & \textbf{3.3$\pm$0.5} & \textbf{0.93$\pm$0.2} & \textbf{3.6$\pm$0.5} & 0.9$\pm$0.2 & \textbf{3.1$\pm$0.4}\\
\hline
\end{tabular}
\end{center}
\label{tab:three_paradigms}
\end{table*}

The asterisks ($^*$) in Figs.~\ref{fig:vesus_spk5}-\ref{fig:wavenet_mixed} denote significantly improved performance between the VCGAN-II (F0+Energy) and the alternate methods. This analysis was conducted via a one-sided t-test for each emotion pair at significance level $p<0.05$. 
We observe that while the three VCGAN models perform similarly, VCGAN-II tends to have more robust performance across evaluation settings. 
The traditional Cycle-GAN does well on the single-speaker evaluation, likely because this architecture was developed for voice conversion and can capitalize on individual speaker characteristics. However, it achieves significantly lower emotional saliency as the evaluation becomes more difficult (i.e., multi-speaker, out-of-speaker, Wavenet). The GMM has variable emotion conversion performance in the single-speaker setting, but fails to generate intelligible speech in the multi-speaker paradigms, and performs poorly in the other two evaluations. Finally, the Bi-LSTM achieve low emotional saliency but consistently high MOS score. This is due to the fact that it collapses into an identity transformation and fails to modify the utterance at all. 
From Table~\ref{tab:three_paradigms}, we conclude that our VCGAN-II models achieve the best trade off between emotional saliency and speech reconstruction quality. 
Thus, combining F0 contour and spectrum modification (via energy) into a single unified framework can achieve much better performance on emotion conversion and reconstruction quality assessment tasks than modeling them separately. 

By using diffeomorphic registration for the F0 and energy contour, our novel framework offers some key advantages over the standard wavelet parameterization. 
Furthermore, our momenta-based approach does not require any speaker/cohort specific normalization to match the range of loudness and fundamental frequency. 
The deformation process takes care of the individual ranges, thereby, allowing the VCGAN to automatically adapt to the test speaker. 
Additionally, the KL divergence penalty between the target data density and the generator estimated density constrains the model to behave in a predictable manner. 
The conditional independence of the target spectrum and target F0 contour (Fig.~\ref{fig:graphical_model}) is another notable aspect of our approach; empirically, it helps preserve the naturalness of the modified speech.

\section{Conclusion}
In this paper, we have proposed a novel method for robust emotion conversion. 
Our technique uses a modified version of Cycle-GAN called variational Cycle-GAN (VCGAN). 
VCGAN was derived as an upper bound on the KL-divergence penalty between the target data distribution and the generator estimated distribution. 
We showed that this led to a new joint density discriminator which constrained the forward-backward generators at the distribution level. 
Empirically, we demonstrated that this distributional matching was better at learning the target densities for emotion conversion. 
In addition, we modeled the features in the target utterance as a smooth warped version of the source. 
This allowed the algorithm to adaptively adjust the F0 and loudness range of a test speaker without any feature normalization. 
We showed that our approach led to a consistent performance across \textcolor{blue}{four} emotion conversion tasks. 
Further, our framework achieved a good balance between the emotion conversion accuracy and the naturalness of synthesized speech, as demonstrated by real-world crowd sourcing experiments. 
We also compared our proposed framework against state-of-the-art emotion conversion baselines from supervised and unsupervised learning domain. Our method universally outperformed these techniques.

\bibliographystyle{IEEEtran}

\bibliography{main_A}



\ifCLASSOPTIONcaptionsoff
  \newpage
\fi

\end{document}


\onecolumn

\title{A Diffeomorphic Flow-based Variational Framework for Multi-speaker Emotion Conversion}

%

\author{Ravi~Shankar,
        Hsi-Wei~Hsieh, 
        Nicolas~Charon, 
        and~Archana~Venkataraman,~\IEEEmembership{Member,~IEEE,}
}

\markboth{Journal of \LaTeX\ Class Files,~Vol.~20, No.~14, August~2021}%
{Shell \MakeLowercase{\textit{et al.}}: Bare Demo of IEEEtran.cls for IEEE Journals}

\maketitle

\appendices

\section{Instability of Cycle Consistency Loss} \label{prop1}

\noindent
\textbf{Proposition 1:} The cycle consistency loss is an upper bound on the difference between the first order moments of the original data distribution and its cyclic counterpart.

\medskip \noindent
\textit{Proof:} This bound can be easily derived using Jensen's inequality and the convexity of $\ell_1$ norm. Namely,
\begin{align} \nonumber
    &E_{x \sim P(X)} \left[\lVert x - G_{\theta}(G_{\gamma}(x)) \rVert_{1}\right] \geq \\
    & \quad \lVert E_{x \sim P(X)} [x - G_{\theta}(G_{\gamma}(x))] \rVert_1 = \lVert \mu_X - \mu_X^c \rVert_1  
\end{align}

\medskip \noindent
\textbf{Proposition 2:} Under mild assumptions, cycle consistency loss is never zero unless the generators are deterministic.

\medskip \noindent
\textit{Proof:} The Cycle-GAN generators learn a transformation function for the input data distribution. 
Let $\hat{X} = G_\theta(G_\gamma(X))$ denote the cyclic transformation by passing through the cascade of forward and backward generators, sequentially. 
To simplify the derivation, we assume that the data is 1-dimensional and $\mathcal{X} = \{x: P_{Data}(x)>0 \}$ denote the support of data distribution. 
The cyclic transformation, therefore, allows us to approximate the conditional distribution $P(\hat{X} | X)$ represented by:
\begin{equation*}
\hat{X} = X + \epsilon \text{ and } \epsilon \sim N(0, \tau^2) \text{ i.e, } P(\hat{X} | X) \sim N(\hat{X}; X, \tau^2)
\end{equation*}
Here, $\tau$ is the variance due to noise in the sampling process. 
This is a reasonable model when the generators can faithfully approximate the data generating distribution. 
Further, the sampling process in GANs is implemented via feeding a Gaussian random vector to the generator. 
Therefore, the expected cyclic consistency loss (after training) can be written as:
\begin{align}
& E\big[ \Arrowvert X - G_\theta(G_\gamma(X)) \Arrowvert_1 \big] = E_{(X,\hat{X})}\big[ \Arrowvert X - \hat{X} \Arrowvert_1 \big] \nonumber \\
& \phantom{E\big[ \Arrowvert X - G_\theta(G_\gamma(X)) \Arrowvert_1 \big]} = \int_{\mathscr{X}} p(x) \Big( \int_{-\infty}^{\infty} \Arrowvert x - \hat{x} \Arrowvert_1 \; p(\hat{x}|x) \; d\hat{x} \Big) \; dx \nonumber \\
& \phantom{E\big[ \Arrowvert X - G_\theta(G_\gamma(X)) \Arrowvert_1 \big]} = \int_{\mathscr{X}} p(x) \Big( \int_{-\infty}^{\infty} \Arrowvert x - \hat{x} \Arrowvert_1 \; \frac{1}{\sqrt{2\pi}\tau} \exp{\frac{-1}{2\tau^2} \Arrowvert x - \hat{x} \Arrowvert_2^2} \; d\hat{x} \Big) \; dx \nonumber \\
& \phantom{E\big[ \Arrowvert X - G_\theta(G_\gamma(X)) \Arrowvert_1 \big]} = \int_{\mathscr{X}} p(x) \Big( \sqrt{\frac{2}{\pi}}\tau \Big) \; dx \qquad \; \; \Big(\text{ using } \int_{-\infty}^{\infty} \frac{\Arrowvert z \Arrowvert_1}{\sqrt{2\pi}} \exp{\frac{-1}{2}\Arrowvert z \Arrowvert_2^2} \; dz  = \sqrt{\frac{2}{\pi}}\Big) \nonumber \\
& \phantom{E\big[ \Arrowvert X - G_\theta(G_\gamma(X)) \Arrowvert_1 \big]} = \sqrt{\frac{2}{\pi}}\tau \qquad \qquad \qquad \qquad \text{ $\Big(\sqrt{\frac{2}{\pi}} n \tau$ when $X, \hat{X} \; \epsilon \;\mathbb{R}^n \Big)$} \nonumber \\
\end{align}
The expected loss depends on the variance of the sampling process obtained by cascading the two generators. 
Thus, the minimum loss is not zero unless the sampling variance vanishes, i.e, the generators become completely deterministic. 
When the data is high-dimensional, the situation is considerably worse even if the sampling variance is small because of the linear dependency on the dimensionality of data. 
Therefore, cyclic gradient from consistency loss can lead to poor model calibration. 

\newpage
\section{Discriminator Objective}
In this section we show how the KL divergence penalty leads to a fundamentally different discriminator which acts on the joint densities instead of the marginals. 
We begin with the proposed KL divergence loss from the main text (equation 7). 
Then, using the expression $P_{\theta}(\mathbf{S}_A, \mathbf{p}_A) = \int \int P_{\theta}(\mathbf{S}_A, \mathbf{p}_A | \mathbf{S}_B, \mathbf{p}_B) \times P(\mathbf{S}_B, \mathbf{p}_B) \; d\mathbf{S}_B \; d\mathbf{p}_B$, we can write:
\vspace{0.05cm}
\begin{align}
& \mathscr{L}_{G_\gamma} = KL\Big(P(\mathbf{S}_A, \mathbf{p}_A) \Arrowvert P_{\theta}(\mathbf{S}_A, \mathbf{p}_A)\Big) \nonumber \\
& \phantom{\mathscr{L}_{G_\gamma}} = \int \int P(\mathbf{S}_A, \mathbf{p}_A) \log \frac{P(\mathbf{S}_A, \mathbf{p}_A)}{P_{\theta}(\mathbf{S}_A, \mathbf{p}_A)} \; d\mathbf{S}_A \; d\mathbf{p}_A \nonumber \\
& \phantom{\mathscr{L}_{G_\gamma}} = \int \int P(\mathbf{S}_A, \mathbf{p}_A) \log \frac{P(\mathbf{S}_A, \mathbf{p}_A)}{\int \int P_{\theta}(\mathbf{S}_A, \mathbf{p}_A | \mathbf{S}_B, \mathbf{p}_B) P(\mathbf{S}_B, \mathbf{p}_B) \; d\mathbf{S}_B \; d\mathbf{p}_B} \; d\mathbf{S}_A \; d\mathbf{p}_A \nonumber \\
& \phantom{\mathscr{L}_{G_\gamma}} = -\int \int P(\mathbf{S}_A, \mathbf{p}_A) \log \Bigg(\int \int P_{\theta}(\mathbf{S}_A, \mathbf{p}_A | \mathbf{S}_B, \mathbf{p}_B) P(\mathbf{S}_B, \mathbf{p}_B) \; d\mathbf{S}_B \; d\mathbf{p}_B \Bigg) \; d\mathbf{S}_A \; d\mathbf{p}_A \nonumber \\
& \phantom{\qquad \qquad \qquad \qquad \qquad \qquad \qquad \qquad} - H(\mathbf{S}_A, \mathbf{p}_A) \nonumber \\
& \phantom{\mathscr{L}_{G_\gamma}} = -\int \int P(\mathbf{S}_A, \mathbf{p}_A) \log \Bigg(\int \int \frac{P_{\theta}(\mathbf{S}_A, \mathbf{p}_A | \mathbf{S}_B, \mathbf{p}_B) P_{\gamma}(\mathbf{S}_B, \mathbf{p}_B | \mathbf{S}_A, \mathbf{p}_A) P(\mathbf{S}_B, \mathbf{p}_B)}{P_{\gamma}(\mathbf{S}_B, \mathbf{p}_B | \mathbf{S}_A, \mathbf{p}_A)} \; d\mathbf{S}_B \; d\mathbf{p}_B \Bigg) \; d\mathbf{S}_A \; d\mathbf{p}_A \nonumber \\
& \phantom{\qquad \qquad \qquad \qquad \qquad \qquad \qquad \qquad} - H(\mathbf{S}_A, \mathbf{p}_A) \nonumber \\
& \phantom{\mathscr{L}_{G_\gamma}} = -\int \int P(\mathbf{S}_A, \mathbf{p}_A) \log \Bigg(\int P_{\gamma}(\mathbf{p}_B | \mathbf{S}_A, \mathbf{p}_A) \int \frac{P_{\theta}(\mathbf{S}_A, \mathbf{p}_A | \mathbf{S}_B, \mathbf{p}_B) P_{\gamma}(\mathbf{S}_B | \mathbf{p}_B,  \mathbf{S}_A, \mathbf{p}_A) P(\mathbf{S}_B, \mathbf{p}_B)}{P_{\gamma}(\mathbf{S}_B,  \mathbf{p}_B | \mathbf{S}_A, \mathbf{p}_A)} \; d\mathbf{S}_B \; d\mathbf{p}_B \Bigg) \; d\mathbf{S}_A \; d\mathbf{p}_A \nonumber \\
& \phantom{\qquad \qquad \qquad \qquad \qquad \qquad \qquad \qquad} - H(\mathbf{S}_A, \mathbf{p}_A) \nonumber \\
& \phantom{\mathscr{L}_{G_\gamma}} \leq -\int \int P(\mathbf{S}_A, \mathbf{p}_A) \Bigg(\int P_{\gamma}(\mathbf{p}_B | \mathbf{S}_A, \mathbf{p}_A) \log \int \frac{P_{\theta}(\mathbf{S}_A, \mathbf{p}_A | \mathbf{S}_B, \mathbf{p}_B) P_{\gamma}(\mathbf{S}_B | \mathbf{p}_B,  \mathbf{S}_A, \mathbf{p}_A) P(\mathbf{S}_B, \mathbf{p}_B)}{P_{\gamma}(\mathbf{S}_B,  \mathbf{p}_B | \mathbf{S}_A, \mathbf{p}_A)} \; d\mathbf{S}_B \; d\mathbf{p}_B \Bigg) \; d\mathbf{S}_A \; d\mathbf{p}_A \nonumber \\
& \phantom{\qquad \qquad \qquad \qquad \qquad \qquad \qquad \qquad} - H(\mathbf{S}_A, \mathbf{p}_A) \nonumber \\
\nonumber
& \phantom{\mathscr{L}_{G_\gamma}} \leq -\int \int P(\mathbf{S}_A, \mathbf{p}_A) \Bigg(\int \int  P_{\gamma}(\mathbf{S}_B, \mathbf{p}_B | \mathbf{S}_A, \mathbf{p}_A) \log \frac{P_{\theta}(\mathbf{S}_A, \mathbf{p}_A | \mathbf{S}_B, \mathbf{p}_B) P(\mathbf{S}_B, \mathbf{p}_B)}{P_{\gamma}(\mathbf{S}_B,  \mathbf{p}_B | \mathbf{S}_A, \mathbf{p}_A)} \; d\mathbf{S}_B \; d\mathbf{p}_B \Bigg) \; d\mathbf{S}_A \; d\mathbf{p}_A \nonumber \\
& \phantom{\qquad \qquad \qquad \qquad \qquad \qquad \qquad \qquad} - H(\mathbf{S}_A, \mathbf{p}_A) \nonumber \\
& \phantom{\mathscr{L}_{G_\gamma}} = -\int \int P(\mathbf{S}_A, \mathbf{p}_A) \Bigg(\int \int  P_{\gamma}(\mathbf{S}_B, \mathbf{p}_B | \mathbf{S}_A, \mathbf{p}_A) \log \frac{P_{\theta}(\mathbf{S}_A, \mathbf{p}_A | \mathbf{S}_B, \mathbf{p}_B) P(\mathbf{S}_B, \mathbf{p}_B) P(\mathbf{S}_A, \mathbf{p}_A)}{P_{\gamma}(\mathbf{S}_B,  \mathbf{p}_B | \mathbf{S}_A, \mathbf{p}_A) P(\mathbf{S}_A, \mathbf{p}_A)} \; d\mathbf{S}_B \; d\mathbf{p}_B \Bigg) \; d\mathbf{S}_A \; d\mathbf{p}_A \nonumber \\
& \phantom{\qquad \qquad \qquad \qquad \qquad \qquad \qquad \qquad} - H(\mathbf{S}_A, \mathbf{p}_A) \nonumber \\
& \phantom{\mathscr{L}_{G_\gamma}} = -\int \int P(\mathbf{S}_A, \mathbf{p}_A) \Bigg(\int \int  P_{\gamma}(\mathbf{S}_B, \mathbf{p}_B | \mathbf{S}_A, \mathbf{p}_A) \log \frac{P_{\theta}(\mathbf{S}_A, \mathbf{p}_A | \mathbf{S}_B, \mathbf{p}_B) P(\mathbf{S}_B, \mathbf{p}_B)}{P_{\gamma}(\mathbf{S}_B,  \mathbf{p}_B | \mathbf{S}_A, \mathbf{p}_A) P(\mathbf{S}_A, \mathbf{p}_A)} \; d\mathbf{S}_B \; d\mathbf{p}_B \Bigg) \; d\mathbf{S}_A \; d\mathbf{p}_A \nonumber \\
& \phantom{\qquad \qquad \qquad \qquad \qquad \qquad \qquad \qquad} - H(\mathbf{S}_A, \mathbf{p}_A) + H(\mathbf{S}_A, \mathbf{p}_A) \nonumber \\
& \phantom{\mathscr{L}_{G_\gamma}} = \int \int P(\mathbf{S}_A, \mathbf{p}_A) \Bigg(\int \int  P_{\gamma}(\mathbf{S}_B, \mathbf{p}_B | \mathbf{S}_A, \mathbf{p}_A) \log \frac{P_{\gamma}(\mathbf{S}_B, \mathbf{p}_B | \mathbf{S}_A, \mathbf{p}_A) P(\mathbf{S}_A, \mathbf{p}_A)}{P_{\theta}(\mathbf{S}_A, \mathbf{p}_A | \mathbf{S}_B, \mathbf{p}_B) P(\mathbf{S}_B, \mathbf{p}_B)} \; d\mathbf{S}_B \; d\mathbf{p}_B \Bigg) \; d\mathbf{S}_A \; d\mathbf{p}_A \nonumber \\
& \phantom{\mathscr{L}_{G_\gamma}} = \int \int P(\mathbf{S}_A, \mathbf{p}_A) \Bigg(\int \int  P_{\gamma}(\mathbf{S}_B, \mathbf{p}_B | \mathbf{S}_A, \mathbf{p}_A) \Bigg( \log \frac{P_{\gamma}(\mathbf{p}_B | \mathbf{S}_A, \mathbf{p}_A) P(\mathbf{p}_A)}{P_{\theta}(\mathbf{p}_A | \mathbf{S}_B, \mathbf{p}_B) P(\mathbf{p}_B)} \nonumber \\
& \phantom{\qquad \qquad \qquad \qquad \qquad \qquad \qquad \qquad \qquad \qquad \qquad} + \log \frac{P_{\gamma}(\mathbf{S}_B | \mathbf{S}_A, \mathbf{p}_B) P(\mathbf{S}_A | \mathbf{p}_A)}{P_{\theta}(\mathbf{S}_A | \mathbf{S}_B, \mathbf{p}_A) P(\mathbf{S}_B | \mathbf{p}_B)} \Bigg) \; d\mathbf{S}_B \; d\mathbf{p}_B \Bigg) \; d\mathbf{S}_A \; d\mathbf{p}_A \nonumber \\
& \phantom{\mathscr{L}_{G_\gamma}} = \int \int P(\mathbf{S}_A, \mathbf{p}_A) \Bigg(\int \int  P_{\gamma}(\mathbf{S}_B, \mathbf{p}_B | \mathbf{S}_A, \mathbf{p}_A) \Bigg( \log \frac{P_{\gamma}(\mathbf{p}_B | \mathbf{S}_A, \mathbf{p}_A) P(\mathbf{p}_A)}{P_{\theta}(\mathbf{p}_A | \mathbf{S}_B, \mathbf{p}_B) P(\mathbf{p}_B)} \nonumber \\
& \phantom{\qquad \qquad \qquad \qquad \qquad \qquad \qquad \qquad \qquad \qquad \qquad} + \log \frac{P_{\gamma}(\mathbf{S}_B | \mathbf{S}_A, \mathbf{p}_B) P(\mathbf{S}_A | \mathbf{p}_A)}{P_{\theta}(\mathbf{S}_A | \mathbf{S}_B, \mathbf{p}_A) P(\mathbf{S}_B | \mathbf{p}_B)} \Bigg) \; d\mathbf{S}_B \; d\mathbf{p}_B \Bigg) \; d\mathbf{S}_A \; d\mathbf{p}_A \nonumber \\
\label{eqn:KL_derivation}
\end{align}
We have applied Jensen's inequality and following conditional independence from the graphical model to derive the bound:
\begin{align}
& \mathbf{S}_B \perp \mathbf{p}_A | \{\mathbf{S}_A, \mathbf{p}_B\} \; \text{ and } \; \mathbf{S}_A \perp \mathbf{p}_B | \{\mathbf{S}_B, \mathbf{p}_A\} \nonumber \\
\label{eqn:conditional_independeces}
\end{align}
We split the overall loss into two components, one each for F0 contour and spectrum, respectively.
\vspace{0.2cm}
\begin{align}
& \mathscr{L}_{G_\gamma} = \mathscr{L}_{G_\gamma}^p + \mathscr{L}_{G_\gamma}^S \nonumber \\
& \mathscr{L}_{G_\gamma}^p = E_{(\mathbf{S}_A, \mathbf{p}_A) \sim P} \Big[ E_{(\mathbf{S}_B, \mathbf{p}_B) \sim P_\gamma} \Big[ \log \frac{P_{\gamma}(\mathbf{p}_B | \mathbf{S}_A, \mathbf{p}_A) P(\mathbf{p}_A)}{P_{\theta}(\mathbf{p}_A | \mathbf{S}_B, \mathbf{p}_B) P(\mathbf{p}_B)} \Big] \Big] \nonumber \\
& \mathscr{L}_{G_\gamma}^S = E_{(\mathbf{S}_A, \mathbf{p}_A) \sim P} \Big[ E_{(\mathbf{S}_B, \mathbf{p}_B) \sim P_\gamma} \Big[ \log \frac{P_{\gamma}(\mathbf{S}_B | \mathbf{S}_A, \mathbf{p}_B) P(\mathbf{S}_A | \mathbf{p}_A)}{P_{\theta}(\mathbf{S}_A | \mathbf{S}_B, \mathbf{p}_A) P(\mathbf{S}_B | \mathbf{p}_B)} \Big] \Big] \nonumber \\
\label{eqn:split_loss}
\end{align}
We approximate the log ratio of densities in $\mathscr{L}_{G_\gamma}^p$ and $\mathscr{L}_{G_\gamma}^S$ by two distinct discriminators. 
Each generator in our VCGAN gets feedback from these two discriminators which are simultaneously trained to differentiate between the pair of distributions. 

\section{Instability of Joint Discriminator}
\begin{figure*}[t]
  \centering
  \includegraphics[width=0.95\textwidth, height=10cm]{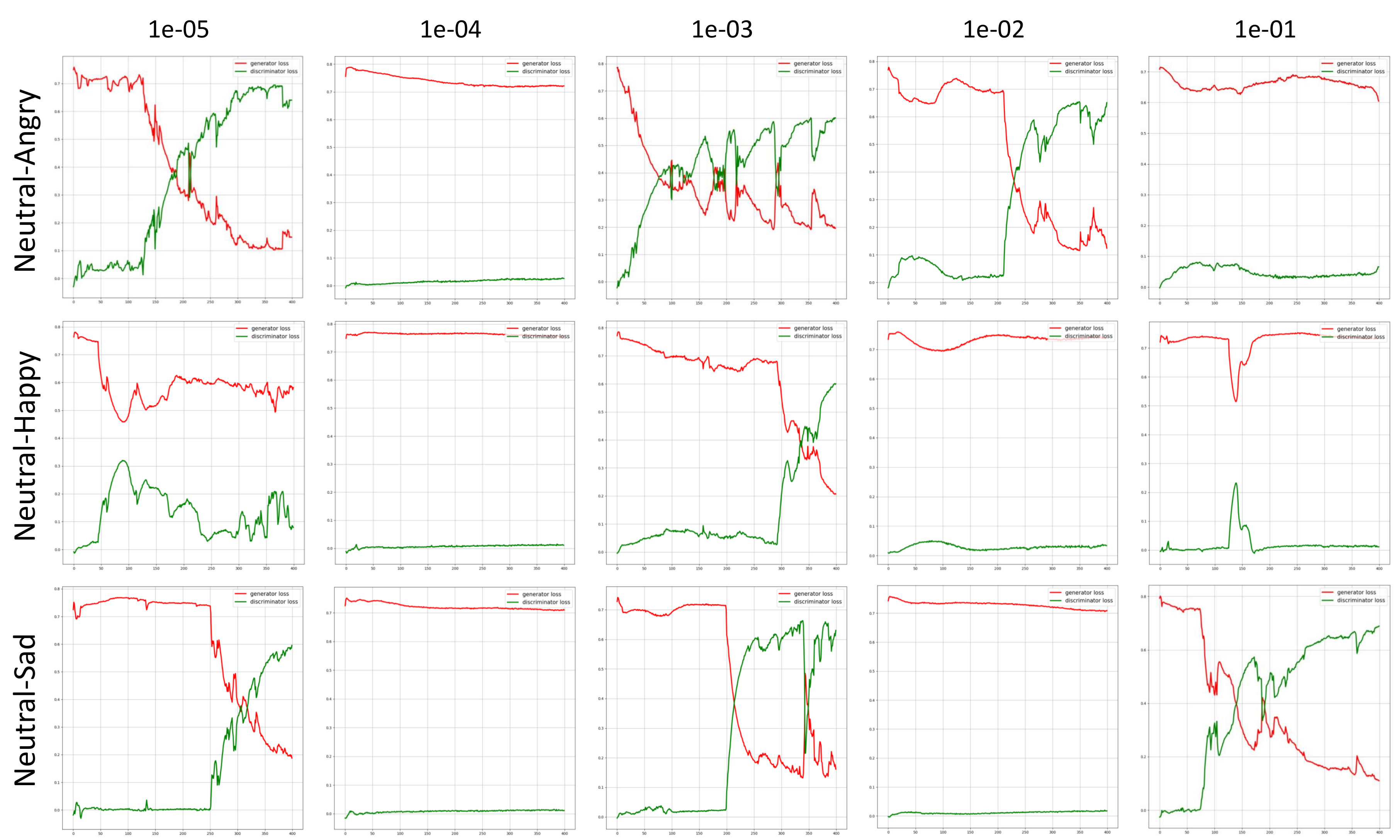}
  \caption{Generator (red) and discriminator (green) losses do not saturate at an equilibrium point when using a joint discriminator. The columns corresponds to different hyper-parameters of $\lambda_{c_1}$ for training neutral-to-angry (top row), neutral-to-happy (middle row) and neutral-to-sad (bottom row). Note the divergence of discriminator and generator loss indicating instability in training the model and in most cases, mode collapse.}
  \label{fig:joint_discriminator}
\end{figure*}
In this section we provide a mathematical description of the GAN optimization in single versus joint discriminator case. We show that the single discriminator loss is not strong enough to effectively backpropagate through the generator pipeline due to the a cascade of matrix-vector products. Further, the warping blocks are implemented as linear transformations that reduce the gradients' magnitude while training. The joint discriminator provides additional stability during the training procedure. We validate our hypothesis by plotting generator-discriminator loss curves for multiple combinations of the consistency loss.

Formally, the generator is composed of two blocks, denoted by $f_\theta$ (F0 contour generation) and $g_\phi$ (energy contour generation). We denote the discriminator by $D$ and derive the generator loss term in adversarial setting. First, consider the simple setting of a single discriminator. For simplicity, we remove other conditioning signals in this derivation. The diffeomorphic structure of the model shows up via a transformation of $f_\theta \rightarrow W_{\sigma_1} f_\theta$ and $g_\phi \rightarrow W_{\sigma_2} g_\phi$ where, the matrices $W_{\sigma_1}$ and $W_{\sigma_2}$ represent the warping parameters for pitch and energy, respectively. Mathematically, the generator loss can be written as follows:
\begin{align}
    & \mathcal{L}_g = E_{z} \Big[\log\Big(1 - D\big(W_{\sigma_2}g_\phi \big(W_{\sigma_1}f_\theta(z) \big) \big)\Big)\Big],
\end{align}
where the random variable~$z$ is assumed to be Gaussian for sampling purposes. The gradient of  the loss term $\mathcal{L}_g$ with respect to the F0 neural network parameters $\theta$ is
\begin{align}
    & \text{\footnotesize $\frac{\partial \mathcal{L}_g}{\partial \theta} = E_{z} \Big[ \Big(1 - D\big(W_{\sigma_2}g_\phi \big(W_{\sigma_1}f_\theta(z) \big) \big)\Big)^{-1} \Big(-D^{\prime}\big(W_{\sigma_2}g_\phi \big(W_{\sigma_1}f_\theta(z) \big) \big)\Big) \Big(W_{\sigma_2}g_{\phi}^{\prime}\big(W_{\sigma_1}f_\theta(z)\big)\Big) W_{\sigma_1}\frac{\partial f_{\theta}(z)}{\partial \theta} \Big] $}
    \label{eqn:unified_loss}
\end{align}
In contrast, we now consider the case of two discriminators, one for the F0 conversion and one for energy conversion. The loss function with respect to $f_\theta$ can be written as:
\begin{align}
    \mathcal{L}_g = E_{z} \Big[\log\Big(1 - D\big(W_{\sigma_1}f_\theta(z)\big)\Big)\Big]
\end{align}
The gradient of this loss with respect to $\theta$ can now be computed as follows:
\begin{align}
    \text{\footnotesize $\frac{\partial \mathcal{L}_g}{\partial \theta} = E_{z} \Big[ \Big(1 - D\big(W_{\sigma_1}f_\theta(z) \big)^{-1}\Big) \Big(-D^{\prime}\big(W_{\sigma_1}f_\theta(z)\big)\Big) W_{\sigma_1}\frac{\partial f_{\theta}(z)}{\partial \theta} \Big] $}
    \label{eqn:joint_loss}
\end{align}
Comparing Eq.~(\ref{eqn:joint_loss}) and Eq.~(\ref{eqn:unified_loss}), we can see that the gradient in the joint case has extra multiplicative terms containing the energy block ($g_\phi$) and its warping parameters ($W_{\sigma_1}$). In practice, these multiplications are cascades of matrix-vector multiplications and therefore reduce the intensity of the gradient signal to effectively train the pitch model $f_\theta$.

Fig.~\ref{fig:joint_discriminator} illustrates the loss plots when training our model using a single discriminator for F0 and energy transformation. We plot the generator (red) and discriminator (green) loss for each emotion pair as we vary the hyper-parameter for the cycle consistency loss over several orders of magnitude (1e-5 to 0.1). 
It can be observed from the plots that, the loss curves never converge at the optimal values for the minimax game, which is 0.5 for both the generator and discriminator. Instead, the losses diverge in almost every single setting, which is likely due to the F0 model not having enough gradient information to learn a transformation. Namely, the error signal feedback from the discriminator is not strong enough to backpropagate through both the energy and F0 generator modules. Splitting up the loss terms guards against this behavior.

\section{Neural Network Architectures}

In this section we detail the network architectures used for each deep learning model. This information is presented as a table of sequential blocks, including dimension and number of parameters.

\bigskip \noindent
\textbf{Bi-LSTM:} The Bi-LSTM model~\cite{lstm_emo_conv} modifies the F0 and spectrum using a 5 layer deep LSTM network. 
It further uses the wavelet decomposition of the prosodic features to control the segmental and supra-segmental nature of prosody. 
The rationale behind using Bi-LSTM is to learn long-range dependency between the frames in both, forward and backward direction and to ensure continuity in prediction. 
The presence of multiple layers allows the model to learn high level representation of voice features. 
The authors further adapted the model for each emotion pair by pre-training it on CMU-ARCTIC~\cite{cmu_arctic} corpus.

\begin{table}[h]
    \centering
    \caption{Summary of the Bi-LSTM architecture}
    \begin{tabular}{ |c|c|c| } 
     \hline
     \textbf{Module} & \textbf{dimensions} & \textbf{$\#$ of parameters} \\
     \hline
     Layer 1 & $4(44\times64 + 64^2 + 64)$ & 27904 \\
     \hline
     Layer 2 & $4(64\times96 + 96^2 + 96)$ & 61824 \\
     \hline
     Layer 3 & $4(96\times128 + 128^2 + 128)$ & 115200 \\
     \hline
     Layer 4 & $4(128\times96 + 96^2 + 96)$ & 86400 \\
     \hline
     Layer 5 & $4(96\times64 + 64^2 + 64)$ & 41216 \\
     \hline
     Dropout & - & - \\
     \hline
     Fully connected & $128\times5\times44 + 1$ & 28161 \\
     \hline
     - & - & 360K \\
     \hline
    \end{tabular}
    \label{tab:LSTM_architecture}
\end{table}

\bigskip \noindent
\textbf{Cycle-GAN:} The authors in~\cite{cyclegan_emo_conv} proposed an unsupervised model for F0 and spectrum transformation using Cycle-GAN~\cite{cycle_gan}. 
This method uses a pair of generator and discriminator network to learn a transformation between two domains. 
The generator network is a fully convolutional model consisting of multiple down-sampling and residual layers to learn better representation of the data. 
The discriminator is another fully convolutional model that provides feedback about estimated data density to the generator. 
The cycle consistency loss is what separates the GAN model from a Cycle-GAN model. 
It constrains the sampling process by tying the generator pair via a cascade of forward and backward transformation. 

\begin{table}[h]
    \centering
    \caption{Summary of the Cycle-GAN architecture (Left: Generator, Right: Discriminator)}
    \begin{minipage}[c]{0.45\textwidth}
    \begin{tabular}{ |c|c|c| } 
         \hline
         \textbf{Module} & \textbf{dimensions} & \textbf{$\#$ of parameters} \\
         \hline
         Convolution & 1x15x64 & 960 \\
         \hline
         Gating & 1x15x64 & 960 \\
         \hline
         Downsample 1 & 1x5x128 & 640 \\
         \hline
         Instance Norm & 1x64x2 & 128 \\
         \hline
         Gating & 1x5x128 & 640 \\
         \hline
         Instance Norm & 1x64x2 & 128 \\
         \hline
         Downsample 2 & 1x5x256 & 1280 \\
         \hline
         Instance Norm & 1x32x2 & 64 \\
         \hline
         Gating & 1x5x256 & 1280 \\
         \hline
         Instance Norm & 1x32x2 & 64 \\
         \hline
         Residual 1 & 4x3x512 & 6144 \\
         \hline
         Residual 2 & 4x3x512 & 6144 \\
         \hline
         Residual 3 & 4x3x512 & 6144 \\
         \hline
         Upsample 1 & 2x2x512 & 2048 \\
         \hline
         Upsample 2 & 2x2x256 & 1024 \\
         \hline
         Dropout & - & - \\
         \hline
         Convolution 2 & 1x15x24 & 360 \\
         \hline
         - & - & 28K \\
         \hline
    \end{tabular}
    \label{tab:CGAN_generator_architecture}
    \end{minipage}
    \begin{minipage}[c]{0.45\textwidth}
    \begin{tabular}{ |c|c|c| } 
         \hline
         \textbf{Module} & \textbf{dimensions} & \textbf{$\#$ of parameters} \\
         \hline
         Convolution 1 & 1x3x3x64 & 576 \\
         \hline
         Gating 1 & 1x3x3x64 & 576 \\
         \hline
         Downsample 1 & 1x3x3x128 & 1152 \\
         \hline
         Instance Norm & 1x64x2 & 128 \\
         \hline
         Gating & 1x3x3x128 & 1152 \\
         \hline
         Instance Norm & 1x64x2 & 128 \\
         \hline
         Downsample 2 & 1x3x3x256 & 2304 \\
         \hline
         Instance Norm & 1x32x2 & 64 \\
         \hline
         Gating & 1x3x3x256 & 2304 \\
         \hline
         Instance Norm & 1x32x2 & 64 \\
         \hline
         Downsample 3 & 1x6x3x512 & 9216 \\
         \hline
         Instance Norm & 1x16x2 & 32 \\
         \hline
         Gating & 1x6x3x512 & 9216 \\
         \hline
         Instance Norm & 1x16x2 & 32 \\
         \hline
         Fully connected & 32x1 & 32 \\
         \hline
         - & - & 26.9K \\
         \hline
    \end{tabular}
    \label{tab:CGAN_discriminator_architecture}
    \end{minipage}
    
\end{table}

\bigskip \noindent
\textbf{VCGAN-I:} The VCGAN-I~\cite{vcgan} model is based on Cycle-GAN and uses a fully convolutional neural network for generator and discriminator. 
This model predicts the F0 contour by using an intermediate latent variable between the pair of emotion classes as it is an important prosodic feature from the standpoint of emotion perception. 
VCGAN-I architecture is very similar to the Cycle-GAN model discussed before, but the discriminator's memory footprint is much lower because of 1-D convolutions that replaces 2-D convolutions. 
Due to the latent variable intermediary, the model is able to estimate the range of F0 values robustly, thereby outperforming the Cycle-GAN technique on unseen speaker experiments. 

\begin{table}[h]
    \centering
    \caption{Summary of the VCGAN-I architecture (Left: Generator, Right: Discriminator)}
    \begin{minipage}{0.45\textwidth}
    \begin{tabular}{ |c|c|c| } 
         \hline
         \textbf{Module} & \textbf{dimensions} & \textbf{$\#$ of parameters} \\
         \hline
         Convolution & 1x15x64 & 960 \\
         \hline
         Gating & 1x15x64 & 960 \\
         \hline
         Downsample 1 & 1x5x128 & 640 \\
         \hline
         Instance Norm & 1x64x2 & 128 \\
         \hline
         Gating & 1x5x128 & 640 \\
         \hline
         Instance Norm & 1x64x2 & 128 \\
         \hline
         Downsample 2 & 1x5x256 & 1280 \\
         \hline
         Instance Norm & 1x32x2 & 64 \\
         \hline
         Gating & 1x5x256 & 1280 \\
         \hline
         Instance Norm & 1x32x2 & 64 \\
         \hline
         Residual 1 & 4x3x512 & 6144 \\
         \hline
         Residual 2 & 4x3x512 & 6144 \\
         \hline
         Residual 3 & 4x3x512 & 6144 \\
         \hline
         Upsample 1 & 2x2x512 & 2048 \\
         \hline
         Upsample 2 & 2x2x256 & 1024 \\
         \hline
         Dropout & - & - \\
         \hline
         Convolution 2 & 1x15x24 & 360 \\
         \hline
         - & - & 28K \\
         \hline
    \end{tabular}
    \end{minipage}
    \begin{minipage}{0.45\textwidth}
    \begin{tabular}{ |c|c|c| } 
         \hline
         \textbf{Module} & \textbf{dimensions} & \textbf{$\#$ of parameters} \\
         \hline
         Convolution 1 & 1x3x64 & 192 \\
         \hline
         Gating 1 & 1x3x64 & 192 \\
         \hline
         Downsample 1 & 1x3x128 & 384 \\
         \hline
         Instance Norm & 1x64x2 & 128 \\
         \hline
         Gating & 1x3x128 & 384 \\
         \hline
         Instance Norm & 1x64x2 & 128 \\
         \hline
         Downsample 2 & 1x3x256 & 768 \\
         \hline
         Instance Norm & 1x32x2 & 64 \\
         \hline
         Gating & 1x3x128 & 384 \\
         \hline
         Instance Norm & 1x32x2 & 64 \\
         \hline
         Downsample 3 & 1x3x256 & 768 \\
         \hline
         Instance Norm & 1x16x2 & 32 \\
         \hline
         Gating & 1x3x256 & 768 \\
         \hline
         Instance Norm & 1x16x2 & 32 \\
         \hline
         Fully connected & 32x1 & 32 \\
         \hline
         - & - & 4.3K \\
         \hline
    \end{tabular}
    \label{tab:VCGAN1_discriminator_architecture}
    \end{minipage}    
\end{table}

\bigskip \noindent
\textbf{VCGAN-II:} VCAGAN-II is what we have proposed in this paper. 
This model is capable of jointly modeling the F0 and energy contour conversion from one emotion to another. 
It is the most lightweight model compared to the baseline models. 
This technique also uses latent variable as an intermediary between the emotion classes and minimizes the KL-divergence between the joint density of F0 and energy of source and target domains. 
Due to the perturbation modeling of F0 and energy, VCGAN-II achieves better F0 and energy prediction as validated in our experiments.  

\begin{table}[h]
    \centering
    \caption{Summary of the VCGAN-II architecture (Left: Generator, Right: Discriminator)}
    \begin{minipage}{0.45\textwidth}
    \begin{tabular}{ |c|c|c| } 
         \hline
         \textbf{Module} & \textbf{dimensions} & \textbf{$\#$ of parameters} \\
         \hline
         Convolution & 1x15x64 & 960 \\
         \hline
         Gating & 1x15x64 & 960 \\
         \hline
         Downsample 1 & 1x5x128 & 640 \\
         \hline
         Instance Norm & 1x64x2 & 128 \\
         \hline
         Gating & 1x5x128 & 640 \\
         \hline
         Instance Norm & 1x64x2 & 128 \\
         \hline
         Downsample 2 & 1x5x128 & 640 \\
         \hline
         Instance Norm & 1x32x2 & 64 \\
         \hline
         Gating & 1x5x128 & 640 \\
         \hline
         Instance Norm & 1x32x2 & 64 \\
         \hline
         Residual 1 & 4x3x512 & 6144 \\
         \hline
         Residual 2 & 4x3x512 & 6144 \\
         \hline
         Upsample 1 & 2x2x512 & 2048 \\
         \hline
         Upsample 2 & 2x2x256 & 1024 \\
         \hline
         Dropout & - & - \\
         \hline
         Convolution 2 & 1x15x1 & 15 \\
         \hline
         - & - & 20.2K \\
         \hline
    \end{tabular}
    \label{tab:VCGAN2_generator_architecture}
    \end{minipage}
    \begin{minipage}{0.45\textwidth}
    \begin{tabular}{ |c|c|c| } 
         \hline
         \textbf{Module} & \textbf{dimensions} & \textbf{$\#$ of parameters} \\
         \hline
         Convolution 1 & 1x3x64 & 192 \\
         \hline
         Gating 1 & 1x3x64 & 192 \\
         \hline
         Downsample 1 & 1x3x128 & 384 \\
         \hline
         Instance Norm & 1x64x2 & 128 \\
         \hline
         Gating & 1x3x128 & 384 \\
         \hline
         Instance Norm & 1x64x2 & 128 \\
         \hline
         Downsample 2 & 1x3x256 & 768 \\
         \hline
         Instance Norm & 1x32x2 & 64 \\
         \hline
         Gating & 1x3x128 & 384 \\
         \hline
         Instance Norm & 1x32x2 & 64 \\
         \hline
         Downsample 3 & 1x3x256 & 768 \\
         \hline
         Instance Norm & 1x16x2 & 32 \\
         \hline
         Gating & 1x3x256 & 768 \\
         \hline
         Instance Norm & 1x16x2 & 32 \\
         \hline
         Fully connected & 32x1 & 32 \\
         \hline
         - & - & 4.3K \\
         \hline
    \end{tabular}
    \label{tab:VCGAN2_discriminator_architecture}
    \end{minipage}
\end{table}


\newpage
\section{Hyperparameters of the VCGAN-II model}

We use a grid search based on the training and validation set of the mixed speaker evaluation to select the model hyperparameters. The generator loss $G_\gamma$ is given by:
\begin{align}
    & \mathscr{L}_{G_{\gamma}} = \lambda_{c_1} E \Big[ \Arrowvert \mathbf{p}_{A} - \mathbf{p}_{A}^{c} \Arrowvert \Big] + \lambda_{m} E \Big[ \Arrowvert \nabla \mathbf{m}_{p} \Arrowvert_2^2 + \Arrowvert \nabla \mathbf{m}_{e} \Arrowvert_2^2 \Big] \nonumber \\
    & \phantom{\;\; L_{G_{\gamma}}} + \lambda_i E \Big[ \Arrowvert \mathbf{e}_A - \mathbf{e}_A^I \Arrowvert \Big] + \lambda_{c_2} E \Big[ \Arrowvert \mathbf{e}_A - \mathbf{e}_A^c \Arrowvert \Big] \nonumber \\
    & \phantom{\;\; L_{G_{\gamma}}} + \lambda_{d} E_{(\mathbf{S}_{A},\mathbf{p}_{A})} \Big[E_{(\mathbf{S}_B, \mathbf{p}_{B}) \sim P_{\gamma}}  \Big[ \log \Big( D_{\gamma}(\mathbf{S}_{A}, \mathbf{p}_{A}, \mathbf{S}_{B}, \mathbf{p}_{B}) \Big) \Big] \Big], \nonumber
    \label{eqn:generator_loss_final}
\end{align}
which gives rise to four hyperparameters $\{\lambda_{c_1},\lambda_{m},\lambda_{i},\lambda_{c_2},\lambda_{d}\}$. Based on the grid search, we select the following configuration

\begin{table}[h]
    \centering
    \caption{Hyperparameters selected for the VCGAN-II Model.}
    \begin{tabular}{ |c|c|c| } 
     \hline
     \textbf{Emotion pair} & \textbf{Hyperparameter} & \textbf{Value} \\ 
     \hline
     \multirow{5}{6em}{Neutral-Angry} & $\lambda_{c_1}$ & 1e-5 \\
     & $\lambda_m$ & 1e-6 \\
     & $\lambda_i$ & 1e-10 \\
     & $\lambda_{c_2}$ & 0.1\\ 
     & $\lambda_d$ & 1\\
     \hline
     \multirow{5}{6em}{Neutral-Happy} & $\lambda_{c_1}$ & 1e-4 \\
     & $\lambda_m$ & 1e-6 \\
     & $\lambda_i$ & 1e-10 \\
     & $\lambda_{c_2}$ & 1e-3\\ 
     & $\lambda_d$ & 1 \\
     \hline
     \multirow{5}{6em}{Neutral-Sad} & $\lambda_{c_1}$ & 1e-4 \\
     & $\lambda_m$ & 1e-6 \\
     & $\lambda_i$ & 1e-10 \\
     & $\lambda_{c_2}$ & 0.1 \\
     & $\lambda_d$ & 1 \\
     \hline
    \end{tabular}
    \label{tab:hyperparams}
\end{table}

We train the generators and discriminators with an initial learning rate of $1e-5$ and $1e-7$, respectively, which empirically produces stable convergence.
The weights are initialized using Xavier's method. 
The minibatch size is fixed at 2 and we train the models for a maximum of 200 epochs in the mixed speaker setting and up to 400 epochs in fold-wise training. 
The models were selected by visual analysis of the generated F0 and energy contour on the validation set. 
The source code can be found at this link: \url{https://github.com/ravi-0841/spect-pitch-gan}. Additionally, the audio samples can be found here: \url{https://engineering.jhu.edu/nsa/wp-content/uploads/2022/07/demo_VCGAN_CGAN.zip}.

\section{Crowd-Sourcing Setup}

\begin{figure*}[t]
  \centering
  \includegraphics[width=0.7\textwidth, height=10cm]{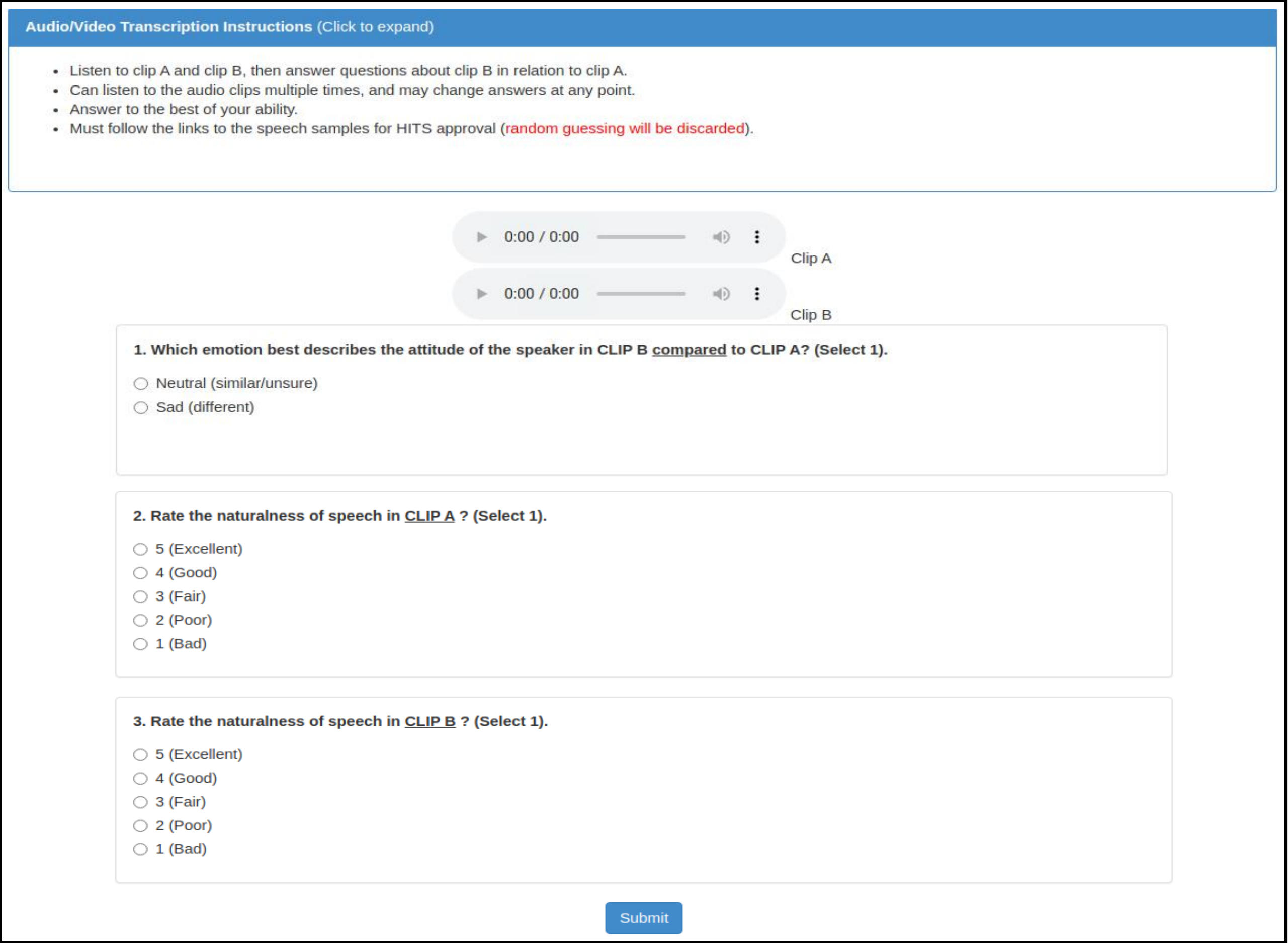}
  \caption{Crowd-sourcing setup used on Amazon Mechanical Turk to collect raters' scores. We ask 3 multiple choice questions to each listener.}
  \label{fig:crowd_sourcing}
\end{figure*}

\begin{table}[h]
    \centering
    \caption{Number of audio samples and ratings collected for each experimental paradigm.}
    \begin{tabular}{ |c|c|c|c|c|c| } 
     \hline
     \textbf{Exp.} & \textbf{Emotion pair} & \textbf{$\#$Unique raters} & \textbf{$\#$Ratings} & \textbf{Max. $\#$ratings/listener} & \textbf{Avg. $\#$ratings/listener}\\ 
     \hline
     \multirow{3}{6em}{Single speaker} & Neutral-Angry & 53 & 300 & 19 & 6 \\
     & Neutral-Happy & 50 & 300 & 19 & 6 \\
     & Neutral-Sad & 48 & 300 & 19 & 6 \\
     \hline
     \multirow{3}{6em}{Mixed speaker} & Neutral-Angry & 42 & 1525 & 65 & 36 \\
     & Neutral-Happy & 42 & 1075 & 65 & 26 \\
     & Neutral-Sad & 45 & 1575 & 63 & 35 \\
     \hline
     \multirow{3}{6em}{Fold-wise} & Neutral-Angry & 54 & 1500 & 70 & 28 \\
     & Neutral-Happy & 57 & 1500 & 65 & 26 \\
     & Neutral-Sad & 58 & 1500 & 79 & 26 \\
     \hline
     \multirow{3}{6em}{Wavenet} & Neutral-Angry & 50 & 1525 & 103 & 30 \\
     & Neutral-Happy & 54 & 1075 & 131 & 20 \\
     & Neutral-Sad & 50 & 1575 & 78 & 32 \\
     \hline
    \end{tabular}
    \label{tab:ratings}
\end{table}

We use a forced choice paradigm for our crowd-sourcing experiment, which is the standard protocol used in emotion conversion papers~\cite{cyclegan_emo_conv, cyclegan_emo_conv_2, lstm_emo_conv}. Namely, we provide both the neutral and emotion converted samples to the listeners, so that there is a baseline reference for each utterance. The forced choice paradigm also mitigates inter-individual variability in both emotional portrayal and emotional perception. 
The listeners (sourced via Amazon Mechanical Turk) are presented with a query of whether or not the converted sample sounds emotional, as compared to the neutral speech. 
We further ask listeners to rate the quality of both the neutral and converted speech utterances.
We collected $5$ ratings for each sample in the test set and compute an average over those ratings to determine emotional saliency. 
Fig.~\ref{fig:crowd_sourcing} illustrates the crowd-sourcing setup a single utterance converted from neutral to sad emotion. 
Table~\ref{tab:ratings} shows the number of unique listeners, total number of rating obtained per emotion class and maximum and average number of ratings provided by an individual listener for each experimental paradigm.

\begin{figure*}[t]
  \centering
  \includegraphics[width=0.7\textwidth, height=8cm]{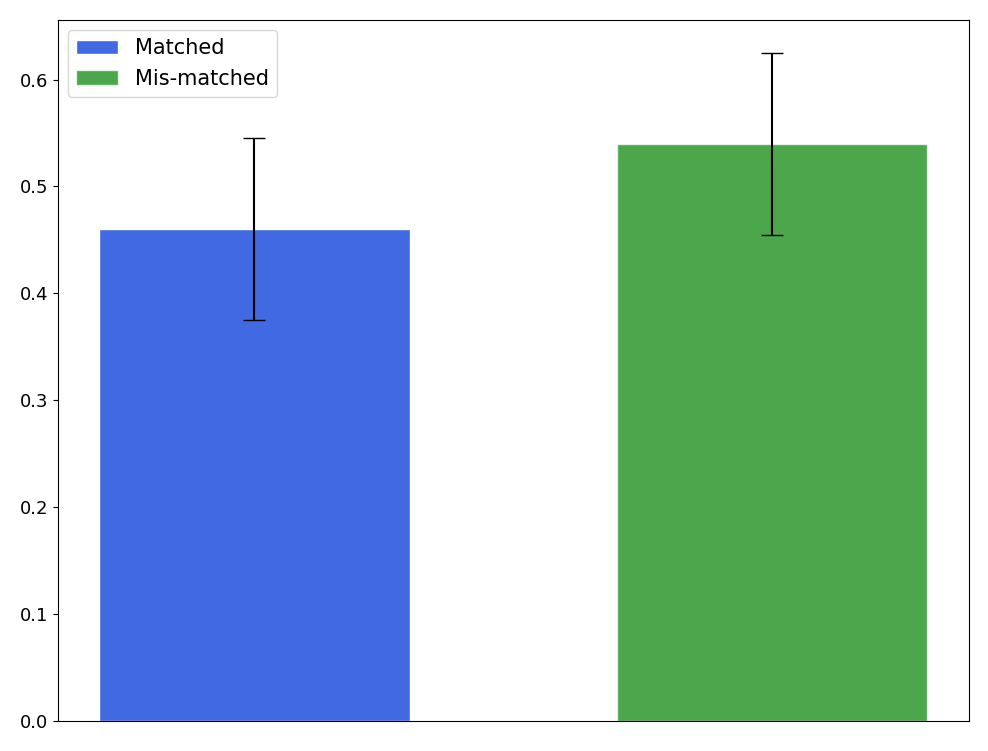}
  \caption{Analysis of F0-Spectrum mismatch by evaluating preference for one audio over another. The results are not significant at level 0.05.}
  \label{fig:mismatch_study}
\end{figure*}

\section{F0-Spectrum Mismatch Analysis}

Our motivation behind modifying just the F0 and energy contours is to avoid degradation in speech quality from directly modifying the spectrum and introducing mismatched harmonics. However, harmonics are also related to pitch, which begs the question of whether modifying the F0 contour will introduce significant distortion to the re-synthesized speech. From a practical standpoint, the WORLD vocoder~\cite{world_vocoder} uses a pitch-synchronous analysis to extract a smoothed spectrum. While the smoothing removes the periodicity introduced by harmonics of the F0, it is still dependent on the fundamental frequency. 
Thus, independent manipulation of the prosodic features could potentially lead to poor speech quality.

To assess this phenomenon, we use WORLD to re-extract a smooth spectrum based on the modified F0 contour and re-synthesize the speech. We then create pairs of audio samples, where each pair consists of the same utterance generated via the re-extracted (matched) spectrum and via our proposed approach that does not modify the spectrum (mismatched). We then conduct a crowd-sourcing experiment that asks the listener to select which utterance in the pair they prefer. 
In this experiment, we generated 10 pairs of audio samples and collected 5 ratings using Amazon Mechanical Turk for each pair. 
The results in Fig.~\ref{fig:mismatch_study} show the preference of raters for the ``matched" versus ``mismatched" audio files. 
Although, there is a slight difference in the means, the preferences are not statistically significantly different.
Thus, we conclude that our strategy of modifying just F0 and energy produces speech of reasonable quality without the need for spectrum modification.

%





\ifCLASSOPTIONcaptionsoff
  \newpage
\fi

\bibliographystyle{IEEEtran}

\bibliography{main_B}